\documentclass[journal,draftclsnofoot,onecolumn,12pt,twoside]{IEEEtranTCOM}
\usepackage{amsmath}
\usepackage{amssymb}
\usepackage{graphicx}
\usepackage[font={small}]{caption}
\usepackage{subcaption}
\usepackage{psfrag}
\usepackage{cite}
\usepackage{multirow}
\usepackage{color}

\newcommand{\bsx}{\boldsymbol{x}}

\newcommand{\bsy}{\boldsymbol{y}}

\newcommand{\bsw}{\boldsymbol{w}}
\newcommand{\bsz}{\boldsymbol{z}}

\newcommand{\bsh}{\boldsymbol{h}}

\newcommand{\bsg}{\boldsymbol{g}}
\newcommand{\bszeta}{\boldsymbol{\zeta}}
\newcommand{\bsv}{\boldsymbol{v}}
\newcommand{\bsI}{\boldsymbol{I}}

\newcommand{\sfT}{\mathsf{T}}
\newcommand{\Ebb}{\mathbb{E}}

\newcommand{\bstheta}{\boldsymbol{\theta}}
\newcommand{\bsTheta}{\boldsymbol{\Theta}}
\newcommand{\bsphi}{\boldsymbol{\phi}}
\newcommand{\bsPhi}{\boldsymbol{\Phi}}
\newcommand{\bspsi}{\boldsymbol{\psi}}
\newcommand{\bsPsi}{\boldsymbol{\Psi}}

\newcommand{\bsD}{\boldsymbol{D}}
\newcommand{\bsO}{\boldsymbol{0}}

\newcommand{\Define}{\stackrel{\Delta}{=}}
\newcommand{\argmax}{\arg\max}
\newcommand{\argmin}{\arg\min}

\newcommand{\diag}{\text{diag}}
\newcommand{\diff}{\mathrm{d}}
\newcommand{\CN}{\mathcal{CN}}

\newcommand{\CircN}{\mathcal{VM}}
\newcommand{\VAR}{\mathsf{VAR}}
\newtheorem{myproposition}{\bf Proposition}
\newtheorem{mycorollary}{\bf Corollary}

\newtheorem{mylemma}{\bf Lemma}
\newtheorem{myremark}{\bf Remark}

\title{ML Detection in Phase Noise Impaired  SIMO Channels with Uplink Training}
\author{Antonios Pitarokoilis,~\IEEEmembership{Student Member, IEEE}, Emil Bj\"{o}rnson,~\IEEEmembership{Member, IEEE}, and Erik G. Larsson,~\IEEEmembership{Senior Member, IEEE}
\thanks{This work was supported by the Swedish Foundation for Strategic Research (SSF) and ELLIIT. The authors are with the Division of Communication Systems, Dept. of Electrical Engineering (ISY), Link\"{o}ping University, 581 83 Link\"{o}ping, Sweden. This work was presented in part at IEEE ICC 2015, London, UK, June 2015.}}

\begin{document}

\IEEEoverridecommandlockouts

\maketitle

\vspace{-15mm}
\begin{abstract}
The problem of maximum likelihood (ML) detection in training-assisted single-input multiple-output (SIMO) systems with phase noise impairments is studied for two different scenarios, i.e. the case when the channel is deterministic and known (constant channel) and the case when the channel is stochastic and unknown (fading channel). Further, two different operations with respect to the phase noise sources are considered, namely, the case of identical phase noise sources and the case of independent phase noise sources over the antennas. In all scenarios the optimal detector is derived for a very general parameterization of the phase noise distribution. Further, a high signal-to-noise-ratio (SNR) analysis is performed to show that symbol-error-rate (SER) floors appear in all cases. The SER floor in the case of identical phase noise sources (for both constant and fading channels) is independent of the number of antenna elements. In contrast, the SER floor in the case of independent phase noise sources is reduced when increasing the number of antenna elements (for both constant and fading channels). Finally, the system model is extended to multiple data channel uses and it is shown that the conclusions are valid for these setups, as well.
\end{abstract}

\section{Introduction}\label{sec:Introduction}

The demand on wireless data services is expected to increase significantly over the next decade. Hence, next generation wireless networks must provide substantially larger data rates. Recently, it has been shown that massive multiple-input multiple-output (\emph{Massive MIMO}) can provide substantial gains in spectral efficiency and radiated energy efficiency \cite{Marzetta10,Rusek13,CommMag13}. In Massive MIMO, $K$ non-cooperative users are served by a base station (BS) with $M$ BS antennas over the same time and frequency resources. When $M$ is significantly larger than $K$ (e.g., one order of magnitude) linear transmit and receive processing techniques are close to optimal and the minimum required radiated power can be reduced as a function of $M$ when a fixed information rate is desired \cite{Hien12TComm,JacobJSAC13}.

In Massive MIMO, the BS uses estimated channel impulse responses to coherently combine the received uplink signals. The quality of the estimated channel state information (CSI) has direct impact on the performance of Massive MIMO systems. Hardware impairments further degrade the acquired channel knowledge. In addition, the deployment of Massive MIMO systems requires the use of inexpensive hardware, so that the monetary cost remains low. Such equipment is likely to have limited accuracy. Hence, the study of the impact of hardware impairments is of particular importance and relevance in Massive MIMO systems. Recently, this area has attracted significant research interest \cite{EmilHardware14}.

An unavoidable hardware impairment in wireless communications is phase noise. Phase noise is introduced in communications systems during the upconversion of the baseband signal to passband and vice versa due to imperfections in the circuitry of local oscillators. Ideally, the local oscillators should produce a sinusoidal wave that is perfectly stable in terms of amplitude, frequency and phase. In the frequency domain that would correspond to a Dirac impulse located at the carrier frequency. However, the phase of the generated carrier of realizable oscillators typically fluctuates. This is manifested by a spectral widening around the carrier frequency in the power spectral density of the local oscillator output. Phase noise can cause significant degradation in scenarios where it varies faster than the channel fading. This happens when the variance of the phase noise innovations and the coherence interval of the channel fading are large \cite[Section IV.C]{EmilTwireless14}. Some scenarios where the phase noise degradation dominates over the degradation due to channel variation are fixed indoor communication and Line-of-Sight (LoS) communication, such as WiFi at millimeter-wave frequencies and wireless broadband-to-home services, respectively. Further, phase noise causes a random rotation of the information signal, i.e. it is a multiplicative distortion. This makes the analysis and mitigation of the phase noise considerably more involved in comparison to additive distortions, such as quantization noise and generic non-linearities. In fact, it appears that Massive MIMO systems are less robust to phase noise than hardware impairments modeled as additive distortions \cite{EmilTwireless14}.

The problem of calculating the capacity of phase noise impaired systems is particularly challenging. Closed-form expressions are not available even for the simplest cases. In \cite{Lapidoth} the author derives the first two terms of the high signal-to-noise--ratio (SNR) expansion of the capacity of a phase noise impaired non-fading single-input single-output (SISO) system for any phase noise process that is ergodic, stationary, and has finite differential entropy rate. In \cite{Durisi12CommL} the first two terms of the high-SNR capacity expansion for the block memoryless phase noise channel are derived. In \cite{DurisiTComm14} a high-SNR capacity upper bound for the Wiener phase noise MIMO channel is derived. Recently, the authors in \cite{DurisiSLOCLO} report approximate upper and lower bounds on the high-SNR capacity for the multiple-input single-output (MISO) and single-input multiple-output (SIMO) phase noise channels and compare the cases where separate and common oscillators are used. Lower bounds on the sum-capacity of multi-user Massive MIMO systems with linear reception and phase noise impairments are recently reported in \cite{EmilTwireless14}, \cite{TWireless} and \cite{Asilomar13}.

The problem of data detection in non-fading channels with phase noise impairments has been extensively studied in the literature. In \cite{Viterbi65TIT} the optimal binary detector for partially coherent channels is derived. Detectors that are optimal in the high-SNR regime are derived in \cite{Foschini73BSTJ}. In \cite{Kam94TComm} the problem of optimal symbol-by-symbol (SBS) detection in SISO systems is investigated when the carrier phase is unknown and it is shown that the computational complexity is prohibitive in the general case. A suboptimal but implementable algorithm is derived when the unknown carrier phase stays constant for a block of consecutive symbols and its performance is compared to the case of exact carrier phase information. In \cite{KhanzadiTCSI} a simulation-based phase noise model is used and the existence EVM floors for SISO systems is shown. The authors of \cite{Colavolpe05JSAC} derive algorithms for SISO phase noise channels without fading based on factor graphs and the sum-product algorithm. An extension of this work for single-user MIMO systems is given in \cite{Rajet13MIMOphaseNoise}, where an estimate of the channel is inserted into the likelihood function as if this estimate were equal to the true channel--resulting in a mismatched detector. A set of soft metrics for the single-user non-fading phase noise channel under various assumptions is derived in \cite{Rajet13SoftMetrics} and their performance is compared. In \cite{Mehrpouyan12} an algorithm for joint data detection and phase noise estimation is derived for single-user MIMO systems and its performance is compared to a derived Cram\'{e}r-Rao bound.

Even though the problem of phase noise in communication systems is extensively studied, there are still many open questions. In particular, the effect of phase noise on beamforming is not fully understood yet. In \cite{Hoehne10} the authors study the effect of phase noise in the \emph{error vector magnitude} (EVM) of an antenna array. They show through analysis and measurements that the EVM at the direction of the main lobe is reduced when independent phase noise sources are used. In \cite{TWireless} achievable rates for the uplink transmission of Massive MIMO systems with time-reversal maximum-ratio-combining in frequency-selective channels are derived for the case of identical and independent phase noise sources. It is observed that the use of independent phase noise sources at the BS results in an increased achievable rate. This result is also supported by a toy example, where the actual capacity can be easily computed and shows that the capacity with independent phase noise sources is larger that the capacity with a single phase noise source. A similar result is reported in \cite{EmilTwireless14}, where the authors show that the phase noise variance can be allowed to increase logarithmically with $M$ when independent phase noise sources are used without losing much in performance. This is not true in the case of identical phase noise sources. Finally, the authors of \cite{DurisiSLOCLO} observe that the \emph{phase noise number} (i.e., the second term in the high-SNR capacity expansion) is higher in the case of independent phase noise sources. The effect of phase noise in linear precoders in downlink Massive MIMO systems is studied recently in \cite{RajetTVT15} and in \cite{EUSIPCO15}. In \cite{EUSIPCO15} the effect of imperfect hardware and the number of oscillators is investigated in distributed Massive MIMO downlink systems and it is shown that superior performance is achieved with independent phase noise sources. 

The results in previous works are not conclusive in the sense that they involve lower or upper bounds on the capacity of the investigated systems. These bounds are useful as the calculation of the exact capacity is an exceedingly complicated problem, however, they do not provide us with a conclusive answer on the fundamental difference between the choice of a single or multiple local oscillators. Motivated by that, we rigorously  derive the optimal detector in phase noise impaired SIMO systems with uplink training for various cases of interest. Our findings explain the effects noted in the prior work; that is, the fact that superior performance was observed in the case of independent phase noise sources. This phenomenon becomes particularly apparent in the high-SNR regime and for a large number of antennas.

We consider a single-user single-input multiple-output (SU-SIMO) system with phase impairments at the multi-antenna receiver. The maximum likelihood (ML) detector is derived under two assumptions on the phase noise processes, namely, when identical (\emph{synchronous} operation) and independent phase noise processes (\emph{non-synchronous} operation) are assumed. The detectors are explicitly given under the assumptions of either constant or fading channels. The phase noise impairments are modeled so that most reasonable distributions of the phase noise increments can be treated in a unified framework. A high-SNR analysis is provided in all the examined scenarios and conclusions are drawn with respect to the symbol-error-rate (SER) performance of the detector when the number of receive antennas, $M$, increases. We observe that in the synchronous operation, an SER floor due to phase noise appears for both assumptions on the channel fading. The SER floor depends on the severity of the phase noise impairments and is independent of the number of receive antennas, $M$. An SER floor at high-SNR appears also in the non-synchronous cases. However, it is shown that this floor can be made arbitrarily small by increasing $M$.

\vspace{-5mm}
\section{System Model}\label{sec:SystemModel}

\begin{figure}[t!]
\psfrag{Pilot}[c][]{\text{Pilot}}
\psfrag{Data}[c][]{\text{Data}}
\psfrag{1}[c][]{\small $1$}
\centering
\includegraphics[height=0.05\textheight]{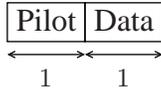}
\caption{Two slot transmission protocol.}\label{fig:SchedulingDiagram}
\end{figure}

In this paper, a single-antenna user communicates with a BS equipped with $M$ antenna elements, which are impaired with phase noise. A simple transmission protocol is considered, which consists of two channel uses (see Fig. \ref{fig:SchedulingDiagram}). In the first channel use a known symbol (pilot) is transmitted and in the second an unknown information symbol is transmitted. Two different cases are treated with respect to the knowledge of the wireless channel. Namely, in the first case, termed as \emph{constant channel (CC)}, the channel is assumed deterministic and known at the receiver \cite{HouShapingGain02},\cite{ICC15}. Hence, the transmitted symbol is observed in the presence of only additive noise and multiplicative phase noise. In the second case, termed as \emph{fading channel (FC)}, the wireless channel is, additionally, assumed unknown at the receiver and Rayleigh fading. We start with the description of the CC for simplicity and subsequently we describe the extension to the FC. 

\subsection{Constant Channel (CC)}\label{sub:ConstantChannel}

During the first channel use, the received complex baseband symbol, $x_m$, at the $m$-th BS antenna is given by
\vspace{-2mm}
\begin{align}\label{eq:ReceiveTrainingAntennam}
x_m=\sqrt{\rho}g_me^{j\theta_m} + w_m,~m=1,\ldots,M,
\end{align}
where $w_m$ is the $m$-th component of the additive white Gaussian noise (AWGN) vector, $\bsw$, distributed as a circularly symmetric complex Gaussian random vector, $\CN(\bsO,\bsI_M)$, $\theta_m$ is the unknown initial phase reference uniformly distributed in the interval $[-\pi,\pi)$. The positive scalar $g_m$ is a known amplitude and $\rho$ is the measured SNR at the $m$-th BS antenna when $g_m=1$. In the second channel use the received symbol at the $m$-th BS antenna is given by
\vspace{-2mm}
\begin{align}\label{eq:ReceiveSymbolAntennam}
y_m=\sqrt{\rho}g_m e^{j(\theta_m+\phi_m)}s + z_m,~m=1,\ldots,M,
\end{align}
where $z_m$ is the $m$-th component of the additive white Gaussian noise vector $\bsz\sim\CN(\bsO,\bsI_M)$, $s$ is the transmitted information symbol selected from a constellation $\mathcal{S}$, such that $\Ebb[s]=0$ and $\Ebb[|s|^2]=1$. The real random variable, $\phi_m$, is a phase noise increment, which is independent of $s$, $\theta_m$, and the AWGN. Since $\phi_m\in[-\pi,\pi)$, a general representation is adopted for the probability density function (pdf), $p_{\Phi_m}(\phi_m)$, of $\phi_m$, based on its Fourier expansion
\begin{align}\label{eq:PhaseNoiseFourier}
p_{\Phi_m}(\phi_m)=\frac{1}{2\pi}\left(\alpha_{m,0}+2\sum_{l=1}^{\infty}\alpha_{m,l}\cos(l\phi_m)\right),
\end{align}
where $\alpha_{m,l},~l=0,1,2,\ldots,$ are known real constants \cite{mardia2009directional,CircularStatisticsBook}. Due to the wrapping of $\phi_m\in[-\pi,\pi)$ the pdf is periodic and the Fourier series expansion exists. The Fourier expansion in \eqref{eq:PhaseNoiseFourier} can represent any pdf in $[-\pi,\pi)$ that is continuous, differentiable, unimodal, even and has zero mean. That includes the circular normal distribution (also known as von Mises or Tikhonov distribution) and the wrapped Gaussian distribution \cite{GoebelIT11}. These two models are predominantly used in the literature to describe phase noise of practical oscillators. For free-running oscillators phase noise is often modeled as a discrete-time Wiener process, where the increments are i.i.d. wrapped Gaussian increments \cite{Demir00}. For oscillators equipped with a PLL, the phase noise increment is well modeled by a random variable from a von Mises (or, equivalently, Tikhonov) distribution \cite{DurisiSLOCLO}.

The setup presented in \eqref{eq:ReceiveTrainingAntennam} and \eqref{eq:ReceiveSymbolAntennam} can model two distinct operations, with respect to phase noise. In the first operation, the random variables $\theta_m$ and $\phi_m$ are independent across the BS antennas (\emph{non-synchronous (NS)} operation). This models a practical distributed antenna deployment where the use of a separate oscillator per BS antenna is required. In the second operation, it holds that $\theta_1\equiv\cdots\equiv\theta_M$ and $\phi_1\equiv\cdots\equiv\phi_M$ (\emph{synchronous (S)} operation). This corresponds to a practical centralized deployment where the same oscillator is used for the downconversion of the received passband signal to the baseband for all BS antennas. Hybrid topologies are also possible, however, they are less interesting to analyze since their SER is expected to lie between the SER of the S and NS operations. Hence, they are not considered in this work. Finally, \eqref{eq:ReceiveTrainingAntennam} and \eqref{eq:ReceiveSymbolAntennam} can be expressed in vector--matrix form as
\begin{align}\label{eq:ConstantChannelMatrixVector}
\bsx&=\sqrt{\rho}\bsTheta\bsg+\bsw,\nonumber\\
\bsy&=\sqrt{\rho}\bsTheta\bsPhi\bsg s+\bsz,
\end{align}
where $\bsx\Define[x_1,\ldots,x_M]^T$, $\bsy\Define[y_1,\ldots,y_M]^T$, $\bsTheta\Define\diag\{e^{j\theta_1},\ldots,e^{j\theta_M}\}$, $\bsPhi\Define\diag\{e^{j\phi_1},\ldots,e^{j\phi_M}\}$, $\bsg\Define[g_1,\ldots,g_M]^T$ is the vector of known amplitudes and $(\cdot)^T$ is the transpose operation.

\subsection{Fading Channel (FC)}\label{sub:FadingChannel}
 
For the FC case, during the first channel use the signal received at the $m$-th BS antenna is given by
\begin{align}\label{eq:ReceiveTrainingAntennamFC}
x_m=\sqrt{\rho}h_m+w_m,
\end{align}
where $h_m$ is the $m$-th component of the zero mean circularly symmetric complex Gaussian channel vector, $\bsh\sim\CN(\bsO,\bsI_M)$. The initial phase $\theta_m$ that is present in \eqref{eq:ReceiveTrainingAntennam} is absorbed in $h_m$. This can be assumed without modifying the statistics of $h_m$, due to the circular symmetry of the channel distribution. During the second channel use, the signal received at the $m$-th BS antenna is given by
\begin{align}\label{eq:ReceiveSymbolAntennamFC}
y_m=\sqrt{\rho}e^{j\phi_m}h_m s+z_m.
\end{align}
The phase noise increment, $\phi_m$, is defined as in \eqref{eq:PhaseNoiseFourier}. The fading coefficients $h_m$ remain constant over both channel uses. Further, the S and NS operations are defined similarly as in Section \ref{sub:ConstantChannel}. Finally, the vector--matrix form of \eqref{eq:ReceiveTrainingAntennamFC} and \eqref{eq:ReceiveSymbolAntennamFC} is given by
\begin{align}\label{eq:FadingChannelMatrixVector}
\bsx&=\sqrt{\rho}\bsh+\bsw,\nonumber\\
\bsy&=\sqrt{\rho}\bsPhi\bsh s+\bsz.
\end{align}

\section{Optimal Detectors}\label{sec:OptimalDetectors}

In this section we describe the maximum likelihood (ML) detector for each of the four different cases described in Section \ref{sec:SystemModel}. The \emph{maximum a posteriori} (MAP) detector is identical to the ML detector, up to an additive constant dependent on the priors of $s$, and hence can be derived by trivial modification of the ML detectors. The optimal detectors are summarized in Table \ref{tab:SummaryOfDetectionRules} for clarity. The BS uses the received vectors $\bsx$ and $\bsy$ jointly, to derive the optimum estimate, $\hat s$, of the transmitted information symbol, $s$, i.e.,
\begin{align}\label{eq:MLDetectorGeneral}
\hat s\Define\argmax_{s\in\mathcal{S}}~~p(\bsx,\bsy|s).
\end{align}
We start with a proposition on the likelihood function for the NS operation.
\begin{myproposition}\label{prop:NSLikelihoodFunction}
The pdf of the received vectors $(\bsx,\bsy)$ given a symbol $s$ for the NS operation is given by
\begin{align}\label{eq:NSLikelihoodFunction}
p(\bsx,\bsy|s)&=A\prod_{m=1}^{M}\left(\beta_{m,0}+2\sum_{l=1}^{\infty}\beta_{m,l}\cos\left(l\zeta_m\right)\right),
\end{align}
where for the CC-NS case we have
\begin{align}
A&=\exp\left(-\|\bsx\|^2-\|\bsy\|^2-\rho (1+|s|^2)\left\|\bsg\right\|^2\right)/\pi^{2M}\label{eq:ADefinitionCC},\\
\beta_{m,l}&=\alpha_{m,l} I_l(2\sqrt{\rho}g_m|s^*y_m|)I_l(2\sqrt{\rho}g_m|x_m|)\label{eq:betaDefinitionCCNS},\\
\zeta_m&=\arg(y_m)-\arg(x_m)-\arg(s)\label{eq:zetaDefinitionNS}
\end{align}
and for the FC-NS case we have
\begin{align}
A&=\frac{e^{-\frac{1+\rho|s|^2}{\rho+\rho|s|^2+1}\|\bsx\|^2-\frac{1+\rho}{\rho+\rho|s|^2+1}\|\bsy\|^2}}{\left(\pi^2\left(\rho+\rho|s|^2+1\right)\right)^M}\label{eq:ADefinitionFC}\\
\beta_{m,l}&=\alpha_{m,l} I_l\left(\frac{2\rho|s^*x_m^*y_m|}{\rho+\rho|s|^2+1}\right)\label{eq:betaDefinitionFCNS}
\end{align}
and $\zeta_m$ as in \eqref{eq:zetaDefinitionNS}. $I_l(\cdot)$ is the $l$-th order modified Bessel function of first kind\cite{VanTreesPartI}, $(\cdot)^*$ is the complex conjugation operation and $\|\cdot\|$ is the Euclidean norm.
\end{myproposition}
\begin{IEEEproof}
See Appendix \ref{app:CCNSLikelihoodFunction} for the CC-NS case and Appendix \ref{app:FCNSLikelihoodFunction} for the FC-NS case.
\end{IEEEproof}
Based on the result in Proposition \ref{prop:NSLikelihoodFunction}, the detector for the NS operation can be derived.
\begin{mycorollary}\label{cor:DetectionRuleNSgeneral}
For a discrete constellation, $\mathcal{S}$, the ML symbol, $\hat s$, for the NS operation
\begin{align}\label{eq:DetectionRuleNSgeneral}
\hat s=\argmax_{s\in\mathcal{S}}~~\mathcal{L}_s^{NS} =\argmax_{s\in\mathcal{S}}~~B + \sum_{m=1}^{M}\ln\left(\beta_{m,0}+2\sum_{l=1}^{\infty}\beta_{m,l}\cos\left(l\zeta_m\right)\right).
\end{align}
where $\mathcal{L}_s^{NS}$ is the decision metric for the symbol $s\in\mathcal{S}$. For the CC-NS case we have
\begin{align}
B&=-\rho |s|^2\left\|\bsg\right\|^2,\label{eq:BDefinitionCC}
\end{align}
$\beta_{m,l}$ as in \eqref{eq:betaDefinitionCCNS} and $\zeta_m$ as in \eqref{eq:zetaDefinitionNS}. For the FC-NS case we have
\begin{align}\label{eq:BDefinitionFC}
B&=-M\ln\left(1+\rho+\rho|s|^2\right)-\frac{1+\rho|s|^2}{1+\rho+\rho|s|^2}\|\bsx\|^2-\frac{1+\rho}{1+\rho+\rho|s|^2}\|\bsy\|^2,
\end{align}
$\beta_{m,l}$ as in \eqref{eq:betaDefinitionFCNS} and $\zeta_m$ as in \eqref{eq:zetaDefinitionNS}.
\end{mycorollary}
\begin{IEEEproof}
The result follows trivially from Proposition \ref{prop:NSLikelihoodFunction} by taking the natural logarithm of \eqref{eq:CCNSLikelihoodFactorization}, $\ln \left(p(\bsx,\bsy|s)\right)$, and dropping the terms that are independent of $s$.
\end{IEEEproof}
The results in Corollary \ref{cor:DetectionRuleNSgeneral} hold for arbitrary constellations. In the following we particularize Corollary \ref{cor:DetectionRuleNSgeneral} for the case of phase shift keying ($N$-PSK) constellations. A rigorous motivation for this choice is deferred to Section \ref{sub:HighSNRSynch}.
\begin{mycorollary}\label{cor:DetectionRuleNSPSK}
For $s$ selected from an $N$-PSK constellation (i.e., $s\in\{e^{j\frac{2\pi n}{N}}\}_{n=0}^{N-1}$) for the NS operation, the ML detection rule is given by \eqref{eq:DetectionRuleNSgeneral} with $B$ as in \eqref{eq:BDefinitionCC} for $|s|^2=1$
\begin{align}\label{eq:DetectionRuleCCNSPSK}
\beta_{m,l}=\alpha_{m,l}I_l\left(2\sqrt{\rho}g_m|x_m|\right)I_l\left(2\sqrt{\rho}g_m|y_m|\right)\text{ and }\zeta_m=\arg(y_m)-\arg(x_m)-\frac{2\pi n}{N}
\end{align}
for the CC-NS case and $B$ as in \eqref{eq:BDefinitionFC} for $|s|^2=1$
\begin{align}\label{eq:DetectionRuleFCNSPSK}
\beta_{m,l}=\alpha_{m,l}I_l\left(\frac{2\rho|x_m^*y_m|}{2\rho+1}\right)\text{ and }\zeta_m=\arg(y_m)-\arg(x_m)-\frac{2\pi n}{N}
\end{align}
for the FC-NS case.
\end{mycorollary}
The counterparts of Proposition \ref{prop:NSLikelihoodFunction} and Corollaries \ref{cor:DetectionRuleNSgeneral} and \ref{cor:DetectionRuleNSPSK} for the S operation are provided in the following.
\begin{myproposition}\label{prop:SLikelihoodFunction}
The pdf of the received vectors $(\bsx,\bsy)$ given a symbol $s$ for the S operation is given by
\begin{align}\label{eq:SLikelihoodFunction}
p(\bsx,\bsy|s)&=A\left(\beta_0+2\sum_{l=1}^{\infty}\beta_l\cos\left(l\zeta\right)\right)
\end{align}
where for the CC-S case we have $A$ as in \eqref{eq:ADefinitionCC},
\begin{align}
\beta_l&=\alpha_l I_l(2\sqrt{\rho}|s^*\bsg^T\bsy|)I_l(2\sqrt{\rho}|\bsg^T\bsx|),\label{eq:betaDefinitionCCS}\\
\zeta&=\arg(\bsg^T\bsy)-\arg(\bsg^T\bsx)-\arg(s)\label{eq:zetaDefinitionCCS}
\end{align}
and for the FC-S case we have $A$ as in \eqref{eq:ADefinitionFC},
\begin{align}
\beta_l=\alpha_l I_l\left(\frac{2\rho|s^*\bsx^H\bsy|}{1+\rho+\rho|s|^2}\right),\label{eq:betaDefinitionFCS}\\
\zeta=\arg(\bsx^H\bsy)-\arg(s)\label{eq:zetaDefinitionFCS}
\end{align}
where $(\cdot)^H$ is the complex conjugation and transposition operation.
\end{myproposition}
\begin{IEEEproof}
See Appendix \ref{app:CCSLikelihoodFunction} for the CC-S case and Appendix \ref{app:FCSLikelihoodFunction} for the FC-S case.
\end{IEEEproof}
The detector for the S operation is given by Corollary \ref{cor:DetectionRuleSgeneral}.
\begin{mycorollary}\label{cor:DetectionRuleSgeneral}
For a discrete constellation, $\mathcal{S}$, the ML symbol, $\hat s$, for the S operation is\vspace{-3mm}
\begin{align}\label{eq:DetectionRuleSgeneral}
\hat s=\argmax_{s\in\mathcal{S}}~~\mathcal{L}_s^{S} =\argmax_{s\in\mathcal{S}}~~B + \ln\left(\beta_0+2\sum_{l=1}^{\infty}\beta_l\cos\left(l\zeta\right)\right).
\end{align}
where $\mathcal{L}_s^{S}$ is the decision metric for the symbol $s\in\mathcal{S}$. For the CC-S case we have $B$ as in \eqref{eq:BDefinitionCC}, $\beta_l$ as in \eqref{eq:betaDefinitionCCS} and $\zeta$ as in \eqref{eq:zetaDefinitionCCS}. For the FC-S case we have $B$ as in \eqref{eq:BDefinitionFC}, $\beta_l$ as in \eqref{eq:betaDefinitionFCS} and $\zeta$ as in \eqref{eq:zetaDefinitionFCS}.
\end{mycorollary}
\begin{IEEEproof}
The proof is done similarly to the proof of Corollary \ref{cor:DetectionRuleNSgeneral}.
\end{IEEEproof}
For an $N$-PSK constellation the detector in Corollary \ref{cor:DetectionRuleSgeneral} can be particularized as in Corollary \ref{cor:DetectionRuleSPSK}.
\begin{mycorollary}\label{cor:DetectionRuleSPSK}
For $s$ selected from an $N$-PSK constellation (i.e. $s\in\{e^{j\frac{2\pi n}{N}}\}_{n=0}^{N-1}$) for the synchronous operation, the ML detection rule is given by \eqref{eq:DetectionRuleSgeneral} with $B$ as in \eqref{eq:BDefinitionCC} for $|s|^2=1$
\begin{align}\label{eq:DetectionRuleCCSPSK}
\beta_l=\alpha_l I_l\left(2\sqrt{\rho}|\bsg^T\bsx|\right)I_l\left(2\sqrt{\rho}|\bsg^T\bsy|\right)\text{ and }\zeta=\arg(\bsg^T\bsy)-\arg(\bsg^T\bsx)-\frac{2\pi n}{N}
\end{align}
for the CC-S case and $B$ as in \eqref{eq:BDefinitionFC} for $|s|^2=1$
\begin{align}\label{eq:DetectionRuleFCSPSK}
\beta_l=\alpha_l I_l\left(\frac{2\rho|\bsx^H\bsy|}{2\rho+1}\right)\text{ and }\zeta=\arg(\bsx^H\bsy)-\frac{2\pi n}{N}
\end{align}
for the FC-S case.
\end{mycorollary}
\begin{table}
\centering
\begin{tabular}{c|c|c|}
\cline{2-3}    & CC & FC \\ 
\hline \multicolumn{1}{|c|}{\multirow{2}{*}{NS}} & Corollary \ref{cor:DetectionRuleNSgeneral} & Corollary \ref{cor:DetectionRuleNSgeneral}\\ \cline{2-3}
\multicolumn{1}{|c|}{\multirow{2}{*}{}}&  \eqref{eq:DetectionRuleNSgeneral}, \eqref{eq:BDefinitionCC}, \eqref{eq:betaDefinitionCCNS}, \eqref{eq:zetaDefinitionNS}  & \eqref{eq:DetectionRuleNSgeneral}, \eqref{eq:BDefinitionFC}, \eqref{eq:betaDefinitionFCNS}, \eqref{eq:zetaDefinitionNS} \\ 
\hline \multicolumn{1}{|c|}{\multirow{2}{*}{S}} & Corollary \ref{cor:DetectionRuleSgeneral} & Corollary \ref{cor:DetectionRuleSgeneral}\\ \cline{2-3}
\multicolumn{1}{|c|}{\multirow{2}{*}{}} & \eqref{eq:DetectionRuleSgeneral}, \eqref{eq:BDefinitionCC}, \eqref{eq:betaDefinitionCCS}, \eqref{eq:zetaDefinitionCCS}  &  \eqref{eq:DetectionRuleSgeneral}, \eqref{eq:BDefinitionFC}, \eqref{eq:betaDefinitionFCS}, \eqref{eq:zetaDefinitionFCS}\\ 
\hline 
\end{tabular}
\caption{Summary of optimal detection rules for all the cases under investigation.}\label{tab:SummaryOfDetectionRules}
\end{table}

\subsection{Implementation of \eqref{eq:DetectionRuleNSgeneral} and \eqref{eq:DetectionRuleSgeneral}}\label{sub:Impementation}

\begin{figure}[t!]
\centering
\begin{minipage}[b]{0.48\linewidth}
\includegraphics[width=\textwidth]{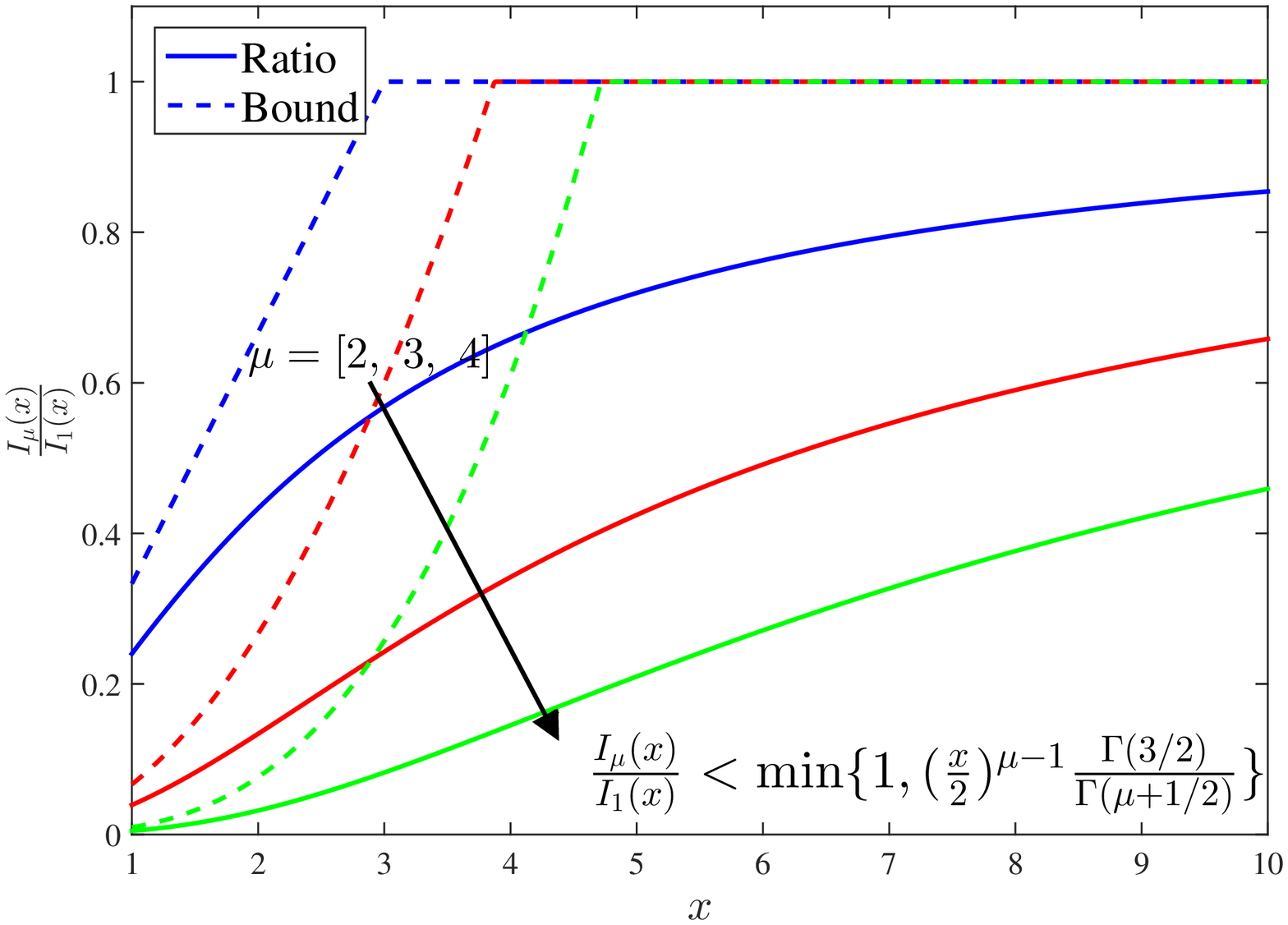}
\caption{Ratio of modified Bessel functions $\frac{I_\mu(x)}{I_1(x)}$ and the bound from Lemma \ref{lem:MBFBound} for $\mu=[2,~3,~4]$ as a function of $x$.}\label{fig:RatioOfBesselIBound}
\end{minipage}
\quad
\begin{minipage}[b]{0.48\linewidth}
\includegraphics[width=\textwidth]{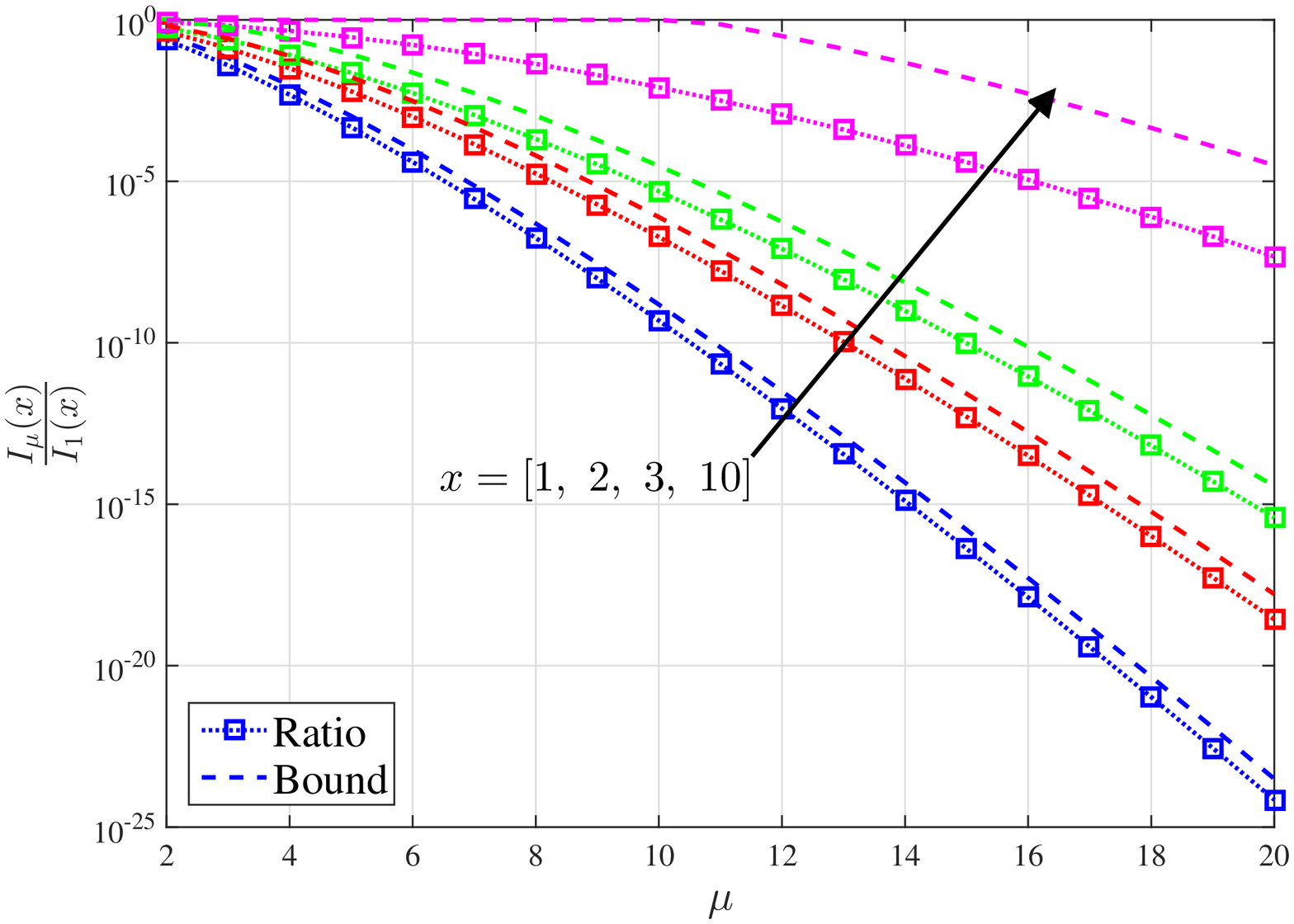}
\caption{Ratio of modified Bessel functions $\frac{I_\mu(x)}{I_1(x)}$ and the bound from Lemma \ref{lem:MBFBound} for a set of fixed values of $x$ as a function of $\mu$.} \label{fig:RatioOfBesselIBoundvaryingMu}
\end{minipage}
\end{figure}

At this point, a comment on computational complexity and implementation issues of the detectors in Corollaries \ref{cor:DetectionRuleNSgeneral} and \ref{cor:DetectionRuleSgeneral} is in order. For the S operation, the calculation of the optimal detectors requires the computation of inner products between vectors of size $M$. For the NS operation, the computational complexity scales also linearly with $M$. Hence, the complexity of the optimal detectors scales in all cases linearly with the number of BS antennas. From \eqref{eq:DetectionRuleNSgeneral} and \eqref{eq:DetectionRuleSgeneral}, the ML detectors involve the calculation of infinite series, where the $l$-th term is a function of the modified Bessel functions, $I_l(\cdot)$. However, the implementation of these detectors is still feasible by appropriate truncation of the infinite series. Since the series converge very fast, the truncation error can be made negligible. The following lemma is useful to demonstrate this claim.
\begin{mylemma}[\!\!\cite{Laforgia91BMF,SIAM1973BoundsMBF}]\label{lem:MBFBound}
For $\mu>\nu>0$ and $x>0$ it holds that
\begin{align}\label{eq:MBFBound}
\frac{I_\nu(x)}{I_\mu(x)}>\max\left\{1,\left(\frac{x}{2}\right)^{\nu-\mu}\frac{\Gamma\left(\mu+\frac{1}{2}\right)}{\Gamma\left(\nu+\frac{1}{2}\right)}\right\},
\end{align}
where $\Gamma(\cdot)$ is the Gamma function $\Gamma(z)\Define\int_{0}^{\infty}e^{-t}t^{z-1}\diff t$, defined for $\Re\{z\}>0$ \cite[8.310]{gradshteyn2007}.
\end{mylemma}
The ratio $\frac{I_\mu(x)}{I_1(x)}$ and the bound in Lemma \ref{lem:MBFBound} are plotted as a function of the argument $x$ for various values of $\mu$ in Fig. \ref{fig:RatioOfBesselIBound}. For all $\mu>1$ and $x$ it holds that $\frac{I_\mu(x)}{I_1(x)}<1$ and for small to moderate values of $x$ we even have $\frac{I_\mu(x)}{I_1(x)}\ll 1$. Further, the ratio $\frac{I_\mu(x)}{I_1(x)}$ is monotonically decreasing in $\mu$ when $x$ is fixed. In fact the rate of decrease is quite fast, since, for any integer $\mu$, $\Gamma(\mu)=(\mu-1)!$ and $x^\mu$ grows at a lesser rate than the factorial function. The rate of decrease as a function of $\mu$ for fixed $x$ is shown in Fig. \ref{fig:RatioOfBesselIBoundvaryingMu}. This fast convergence establishes the fact that only a few terms of the infinite sums are required so that the approximation error is negligible. 

\subsection{The ML Detector for von Mises Phase Noise Increments}\label{sub:vonMisesDetector}

We conclude the section by particularizing the results in a special, but important, case, where the optimal detector structure does not require the implementation of an infinite series. This involves the FC-NS and FC-S cases.
\begin{myproposition}\label{prop:vonMisesFCDetector}
For the fading case with $\phi_m\sim\CircN(0,\kappa)$, distributed as zero mean von Mises random variables with concentration parameter $\kappa$, the likelihood functions can be expressed for the NS operation as
\begin{align}\label{eq:vonMisesFCNSDetector}
p(\bsx,\bsy|s)=A\prod_{m=1}^{M}\frac{I_0\left(\sqrt{\kappa^2+\left(\mathsf{b}_m^{NS}\right)^2+2\kappa \left(\mathsf{b}_m^{NS}\right)\cos\left(\zeta_m\right)}\right)}{I_0(\kappa)}
\end{align}
and for the S operation as
\begin{align}\label{eq:vonMisesFCSDetector}
p(\bsx,\bsy|s)=A\frac{I_0\left(\sqrt{\kappa^2+\left(\mathsf{b}^S\right)^2+2\kappa \left(\mathsf{b}^S\right)\cos\left(\zeta\right)}\right)}{I_0(\kappa)},
\end{align}
where $A$ is defined as in \eqref{eq:ADefinitionFC}, $\mathsf{b}_m^{NS}=\frac{2\rho|s^*x_m^*y_m|}{\rho+\rho|s|^2+1}$, $\mathsf{b}^S=\frac{2\rho|s^*\bsx^H\bsy|}{\rho+\rho|s|^2+1}$, $\zeta_m$ as in \eqref{eq:zetaDefinitionNS} and $\zeta$ as in \eqref{eq:zetaDefinitionFCS}.
\end{myproposition}
\begin{IEEEproof}
See Appendix \ref{app:vonMisesFCDetector}.
\end{IEEEproof}

\section{High SNR Analysis}\label{sec:HighSNR}

\subsection{High SNR Analysis for the Synchronous Operation}\label{sub:HighSNRSynch}

The expressions in Propositions \ref{prop:NSLikelihoodFunction}, \ref{prop:SLikelihoodFunction} and Corollaries \ref{cor:DetectionRuleNSgeneral}-\ref{cor:DetectionRuleSPSK} can be easily implemented, however, it is hard to extract insight on the fundamental behavior of the different operations. Therefore, in this section we present an asymptotic analysis as $\rho\rightarrow\infty$ (high-SNR) for the system models in \eqref{eq:ConstantChannelMatrixVector} and \eqref{eq:FadingChannelMatrixVector} in order to reveal the high-SNR behavior of the detectors. We start with the synchronous operation as it appears to be simpler.
The system model for the CC-S as $\rho\rightarrow\infty$ can be expressed as
\begin{align}\label{eq:highSNRsynch}
\tilde\bsy\Define\lim_{\rho\rightarrow\infty}\frac{\bsy}{\sqrt{\rho}}=e^{j(\theta+\phi)}s\bsg\Rightarrow\tilde\bsy&=\tilde\bsx e^{j\phi}s\Rightarrow\!\begin{cases}
|\tilde\bsx^H\tilde\bsy|=\left\|\bsg\right\|^2|s|\\
\psi=\arg(s)+\phi 
\end{cases}
\end{align}
where $\tilde\bsx\Define\lim_{\rho\rightarrow\infty}\frac{\bsx}{\sqrt{\rho}}=e^{j\theta}\bsg$ and $\psi\Define\arg\left(\tilde\bsx^H\tilde\bsy\right)$. From \eqref{eq:highSNRsynch} it is apparent that the amplitude of $s$ can be decoded error-free. Hence, distinguishing two symbols that have different amplitude is trivial at high-SNR. However, distinguishing two symbols that have different phase is much more challenging. Therefore, in the following, we restrict the study to constellations that encode information only in the phase, that is PSK. Two-dimensional constellations, such as 16-QAM or 64-QAM, can be treated as consisting of multiple disjoint sub-constellations, where the information is conveyed only on the phase as explained in the following. Points of the two dimensional constellation that are at different distances from the origin belong to different sub-constellations with different radii and hence are easily distinguishable at high-SNR. Constellation points that are at the same radius belong to the same sub-constellation and their pairwise error probability can be calculated in a manner similar to PSK constellations. The observation $\psi$ in \eqref{eq:highSNRsynch} is in this case sufficient statistics for the detection of $\arg(s)$. We proceed by deriving the asymptotic SER at high-SNR for PSK constellations.
\begin{myproposition}\label{prop:CCShighSNRErrorFloor}
For the CC-S case, let $\phi$ be zero mean random variable with pdf $p_\Phi(\phi)$ as in \eqref{eq:PhaseNoiseFourier}, where the distribution is unimodal and symmetric around the mean.\footnote{This assumption is not restrictive in practice as the widely accepted models on the phase noise impairments, such as the Wiener model, satisfy it.} Define the error event $\epsilon=\left\{\arg(\hat s)\neq 0\middle|\arg(s)=0\right\}$. Then the SER floor at the high-SNR for equiprobable $N$-PSK symbols is given by
\begin{align}\label{eq:CCShighSNRErrorFloor}
\Pr\left\{\epsilon\right\}&\Define 1-\int_{-\frac{\pi}{N}}^{\frac{\pi}{N}}p_\Phi(\phi)\diff\phi=1-\frac{\alpha_0}{N}-\sum_{l=1}^{\infty}\frac{2\alpha_l}{\pi l}\sin\left(l\frac{\pi}{N}\right).
\end{align}
\end{myproposition}
\begin{mycorollary}\label{cor:highSNRSynch}
From \eqref{eq:CCShighSNRErrorFloor} we observe that there is a non-zero SER floor for the CC-S case, which depends only on the statistics of the phase noise increment and the PSK constellation density, $N$, but is independent of the number of receive antennas, $M$.
\end{mycorollary}
\begin{myremark}\label{rem:EquivalenceCCFCSynch}
The preceding analysis is also true for FC-S by defining $\tilde{\bsx}\Define\bsh$ and $\tilde\bsy\Define\bsh e^{j\phi}s$. Observe that the statistic corresponding to the amplitude \eqref{eq:highSNRsynch} is now $|\tilde\bsx^H\tilde{\bsy}|=\|\bsh\|^2|s|$, where $\bsh$ is stochastic. This generally has an impact of the performance in FC-S, however, in the high-SNR regime the effect of the phase noise increment dominates and the analysis for CC-S applies in FC-S, as well. The performance of FC-S approaches that of CC-S as $M\rightarrow\infty$ due to the hardening of $\frac{1}{M}\|\bsh\|^2\xrightarrow{M\rightarrow\infty}\Ebb\left[|h_1|^2\right]$ almost surely \cite{HochwaldChannelHardening04}.
\end{myremark}

\subsection{High SNR Analysis for the CC-NS case}\label{sub:HighSNRCCNS}

For the CC-NS case based on the model in \eqref{eq:ReceiveTrainingAntennam} and \eqref{eq:ReceiveSymbolAntennam} we define $\tilde x_m\Define\lim_{\rho\rightarrow\infty}\frac{x_m}{\sqrt{\rho}}=e^{j\theta_m}$ and $\tilde y_m\Define\lim_{\rho\rightarrow\infty}\frac{y_m}{\sqrt{\rho}}$. By proceeding similarly to Section \ref{sub:HighSNRSynch} it can be shown that the amplitude can be decoded error free so we concentrate on the transmitted phase, which for the $m$-th BS antenna element is given by $\psi_m\Define\arg(\tilde x_m^*\tilde y_m) = \phi_m + \arg(s)$. The general parameterization of the phase noise increment in \eqref{eq:PhaseNoiseFourier} does not yield mathematically tractable expressions. To get a mathematically tractable expression, we consider that the increments $\phi_m$ are independent $\CircN(0,\kappa)$ random variables for $m=1,\ldots,M$. As noted earlier, the von Mises distribution is used to describe phase noise resulting from oscillators that are equipped with PLL. It is symmetric around its mean, unimodal and is representative of the anticipated statistical behavior of phase noise. Hence, we expect that the  general insights drawn from this choice will hold for other similar distributions. The likelihood function of the observed phases $\bspsi\Define[\psi_1,\ldots,\psi_M]^T$, given the transmitted phase $\arg(s)$, can be expressed as
\begin{align*}
p_{\bsPsi|S}(\bspsi|s)=\prod_{m=1}^{M}p_{\Psi_m|s}(\psi_m|s)=\frac{e^{\kappa\sum_{m=1}^{M}\cos\left(\psi_m-\arg(s)\right)}}{\left(2\pi I_0(\kappa)\right)^M}
\end{align*}
and the corresponding ML decision rule for symbols selected from some alphabet $\mathcal{S}$ is given by
\begin{align}\label{eq:highSNRCCNSdecisionRule}
\hat s=\argmax_{s\in\mathcal{S}}~p_{\bsPsi|S}(\bspsi|s)=\argmax_{s\in\mathcal{S}}~\sum_{m=1}^{M}\cos\left(\psi_m-\arg(s)\right).
\end{align}
\begin{myproposition}\label{prop:highSNRLLRNonSynch}
The decision metric, $\mu_n$, for the symbol $e^{j\frac{2\pi n}{N}}$ from an $N$-PSK constellation based on \eqref{eq:highSNRCCNSdecisionRule} is given by
\begin{align}\label{eq:highSNRLLRNonSynch}
\mu_n&\Define\sum_{m=1}^{M}\left(\cos\left(\psi_m-\frac{2\pi n}{N}\right)-\cos\left(\psi_m\right)\right)=\sum_{m=1}^{M}\sin\left(\frac{\pi n}{N}\right)\sin\left(\psi_m-\frac{\pi n}{N}\right).
\end{align}
If we denote by $\epsilon$ the error event, i.e., the case where the detected symbol $\hat s$ is different from the transmitted symbol $s$, then the SER for equiprobable input symbols is given by
\begin{align}\label{eq:highSNRSERNonSynch}
\Pr\left\{\epsilon\right\}&= \Pr\left\{\bigcup_{n=1}^{N-1}\left\{\mu_n>0\right\}\middle |\arg(s)=0\right\}.
\end{align}
\end{myproposition}

The exact calculation of the probability of error in \eqref{eq:highSNRSERNonSynch} appears formidable. We therefore derive an upper bound on the pairwise symbol error probability of erroneously detecting $s_n=\exp\left(j\frac{2\pi n}{N}\right),~n=1,\ldots,N-1$ when $s_0=1$ was sent. We note that due to the symmetry of the von Mises distribution around its mean and the uniform priors on the input symbols, the conditioning on any particular input symbol does not affect the result. The symbol $s_0=1$ with $\arg(s_0)=0$ is selected for convenience. From \eqref{eq:highSNRLLRNonSynch} the $\mu_n$ is a sum of bounded independent and identically distributed (i.i.d.) random variables $Z_m\Define\sin\left(\frac{\pi n}{N}\right)\sin\left(\psi_m-\frac{\pi n}{N}\right)$. Hence the following lemma can be used.
\begin{mylemma}[\emph{Bernstein Inequality}\cite{Bernstein}]\label{lem:Bernstein}
Let $X_m,~m=1,\ldots,M$, be i.i.d. random variables with $\Ebb [X_m]=0$, $|X_m|<C$ almost surely for some bounded constant $C$, $X_s\Define\sum_{m=1}^{M}X_m$ and $\varsigma\Define\sqrt{\VAR(X_s)}$. Then for all $t>0$\vspace{-4mm}
\begin{align*}
\Pr\left\{X_s>t\varsigma\right\}\leq\exp\left(-\frac{t^2}{2+\frac{2}{3}\frac{C}{\varsigma}t}\right).
\end{align*}
\end{mylemma}
\begin{myproposition}\label{prop:Bernstein}
The pairwise error probability for the detected symbol $\hat s_n$ to be $s_n=\exp\left(j\frac{2\pi n}{N}\right)$, $n=1,\ldots,N-1$, given that the symbol $s_0=1$ was sent is upper bounded by
\begin{align}\label{eq:BernsteinBound}
\Pr\left\{\mu_n>0\middle|s=1\right\}\leq\exp\left(-\frac{M\left(\frac{\sin^2\left(\frac{\pi n}{N}\right)}{\sqrt{\VAR(X_{m,n})}}\frac{I_1(\kappa)}{I_0(\kappa)}\right)^2}{2+\frac{2}{3}\frac{C\sin^2\left(\frac{\pi n}{N}\right)}{\VAR(X_{m,n})}\frac{I_1(\kappa)}{I_0(\kappa)}}\right),
\end{align}
where $C\Define\sin\left(\frac{\pi n}{N}\right)+\sin^2\left(\frac{\pi n}{N}\right)\frac{I_1(\kappa)}{I_0(\kappa)}$ and
\begin{align}\label{eq:VarianceXmn}
\VAR(X_{m,n})=\sin^2\left(\frac{\pi n}{N}\right)\left(\frac{I_1(\kappa)\cos\left(\frac{2\pi n}{N}\right)}{\kappa I_0(\kappa)}+\sin^2\left(\frac{\pi n}{N}\right)\left(1-\frac{I_1^2(\kappa)}{I_0^2(\kappa)}\right)\right).
\end{align}
\end{myproposition}
\begin{IEEEproof}
Let $X_{m,n}\Define\sin\left(\frac{\pi n}{N}\right)\sin\left(\psi_m-\frac{\pi n}{N}\right)+\sin^2\left(\frac{\pi n}{N}\right)\frac{I_1(\kappa)}{I_0(\kappa)}$.
Then $\Ebb [X_{m,n}]=0$,\\ $|X_{m,n}|\leq C\Define\sin\left(\frac{\pi n}{N}\right)+\sin^2\left(\frac{\pi n}{N}\right)\frac{I_1(\kappa)}{I_0(\kappa)}$ and $\VAR(X_{m,n})$ is given by \eqref{eq:VarianceXmn}. Then\small
\begin{align*}
\Pr\left\{\mu_n>0\middle|s=1\right\}&=\Pr\left\{\sum_{m=1}^{M}\sin\left(\frac{\pi n}{N}\right)\sin\left(\psi_m-\frac{\pi n}{N}\right)>0\right\}=\Pr\left\{\sum_{m=1}^{M}X_{m,n}>M\sin^2\left(\frac{\pi n}{N}\right)\frac{I_1(\kappa)}{I_0(\kappa)}\right\}.
\end{align*}
\normalsize Define $\varsigma\Define\sqrt{M}\sqrt{\VAR(X_{m,n})}$ and $t\Define\sqrt{M}\frac{\sin^2\left(\frac{\pi n}{N}\right)}{\sqrt{\VAR(X_{m,n})}}\frac{I_1(\kappa)}{I_0(\kappa)}$.
The result in \eqref{eq:BernsteinBound} follows by application of Lemma \ref{lem:Bernstein}.
\end{IEEEproof}

Proposition \ref{prop:Bernstein} shows that the pairwise error probability for the CC-NS can be reduced exponentially with $M$. In addition, due to the union bound, the SER floor in \eqref{eq:highSNRSERNonSynch} can be reduced arbitrarily to zero by increasing $M$.

\subsection{High SNR Analysis for the FC-NS Case}\label{sub:highSNRFCNS}

Similarly to Remark \ref{rem:EquivalenceCCFCSynch}, for the FC-NS case the following variables are defined $\tilde x_m\Define\lim_{\rho\rightarrow\infty}\frac{x_m}{\sqrt{\rho}}$ and $\tilde y_m\Define\lim_{\rho\rightarrow\infty}\frac{y_m}{\sqrt{\rho}}$, which in the high-SNR regime simplify to $\tilde x_m=h_m$ and $\tilde y_m=e^{j\phi_m}h_m s$. The observation vector $\bsv=[v_1,\ldots,v_m,\ldots,v_M]^T$ with $v_m\Define\tilde x_m^*\tilde y_m=|h_m|^2e^{j\phi_m}s$ is further defined. The derivation of the optimal detection rule even in this regime appears to be mathematically intractable. We seek to find an upper bound on the high-SNR SER floor by using a heuristic suboptimal decision statistic. We assume that $\phi_m$ are i.i.d. $\CircN(0,\kappa)$ for some $\kappa>0$ random variables and we define the decision vector $\bszeta\Define[\zeta_1~\zeta_2]^T$, where
\begin{align}\label{eq:FCNSsuboptimalDetectionRule}
\left[\begin{array}{c}\zeta_1 \\ \zeta_2
\end{array}\right]&\Define\frac{1}{M}\left[\begin{array}{c}
\Re\left\{\sum_{m=1}^{M}v_m\right\}\\ \Im\left\{\sum_{m=1}^{M}v_m\right\}
\end{array}\right]=\frac{1}{M}|s|^2\left[\begin{array}{c}
\sum_{m=1}^{M}|h_m|^2\cos\left(\phi_m+\arg(s)\right)\\ \sum_{m=1}^{M}|h_m|^2\sin\left(\phi_m+\arg(s)\right)
\end{array}\right].
\end{align}
The suboptimal decision rule that we use is the minimum Euclidean distance from a scaled $N$-PSK, i.e.,
\vspace{-5mm}
\begin{align}\label{eq:minimumDistanceFCNS}
\hat s&=\argmin_{s\in\mathcal{S}} \left\|\bszeta-s\right\|,
\end{align}
where $\mathcal{S}$ is the $N$-PSK alphabet. Conditioned on the symbol $s_0=[1~0]^T$ being sent and given the decision rule \eqref{eq:minimumDistanceFCNS}, an error occurs when $\epsilon=\{\hat s\neq s_0|s=s_0\}$. Hence, the symbol error probability can be upper bounded by the union bound as follows:
\begin{align}\label{eq:SERfloorFCNS}
\Pr\left\{\epsilon\right\}&\leq\sum_{n=1}^{N-1}\Pr\left\{\hat s=s_n\middle|s=s_0\right\} 
\end{align}
where $s_n=[\cos(\frac{2\pi n}{N})~\sin\left(\frac{2\pi n}{N}\right)]^T$. 
We provide here a lemma that is useful to bound \eqref{eq:SERfloorFCNS}.
\begin{mylemma}[\emph{Chebychev's Inequality}\cite{Billingsley}]\label{lem:ChebychevsInequality}
Let $X$ be a random variable with mean $\mu$ and variance $\sigma^2$. For any $\varepsilon>0$,\vspace{-6mm}
\begin{align}\label{eq:ChebychevsInequality}
\Pr\left\{\left|X-\mu\right|\geq \varepsilon\right\}\leq\frac{\sigma^2}{\varepsilon^2}.
\end{align}
\end{mylemma}
\begin{mycorollary}\label{cor:highSNRboundFCNS}
The SER floor at high-SNR for the FC-NS case in \eqref{eq:SERfloorFCNS} scales at least as  $O\left(\frac{1}{M}\right)$.
\end{mycorollary}
\begin{IEEEproof}
The pairwise error probabilities, $\Pr\left\{\hat s=s_n\middle|s=s_0\right\}$, are bounded as follows
\begin{align*}
\Pr&\left\{\hat s=s_n\middle|s=s_0\right\}=\Pr\left\{\left\|s_0-\bszeta\right\|>\left\|s_n-\bszeta\right\|\middle|s=s_0\right\}=\Pr\left\{\bszeta^T\left(s_0-s_n\right)<0\middle|s=s_0\right\}\\
&=\Pr\left\{\xi_n<0\right\}=\Pr\left\{-\xi_n+\Ebb[\xi_n]>\Ebb[\xi_n]\right\}\stackrel{(a)}{\leq}\Pr\left\{\left|\xi_n-\Ebb[\xi_n]\right|>\Ebb[\xi_n]\right\}\stackrel{(b)}{\leq}\frac{\VAR(\xi_n)}{\left(\Ebb[\xi_n]\right)^2}
\end{align*}
where $\xi_n\Define\frac{1-\cos\left(\frac{2\pi n}{N}\right)}{M}\sum_{m=1}^{M}|h_m|^2\cos(\phi_m)+\frac{\sin\left(\frac{2\pi n}{N}\right)}{M}\sum_{m=1}^{M}|h_m|^2\sin(\phi_m)$ and\\ $\Ebb[\xi_n]=\left(1-\cos\left(\frac{2\pi n}{N}\right)\right)\frac{I_1(\kappa)}{I_0(\kappa)}$. The inequality in $(a)$ follows from the fact that $\{-\xi_n+\Ebb[\xi_n]>\Ebb[\xi_n]\}\subseteq\{|\xi_n-\Ebb[\xi_n]|>\Ebb[\xi_n]\}$ and $(b)$ follows from Lemma \ref{lem:ChebychevsInequality}. Calculation of the variance $\VAR(\xi_n)$ gives
\begin{align*}
\VAR(\xi_n)=\frac{\left(1-\cos\left(\frac{2\pi n}{N}\right)\right)^2}{M}\left(1+\frac{I_2(\kappa)}{I_0(\kappa)}-\left(\frac{I_1(\kappa)}{I_0(\kappa)}\right)^2\right)+\frac{\left(\sin\left(\frac{2\pi n}{N}\right)\right)^2}{M}\left(1-\frac{I_2(\kappa)}{I_0(\kappa)}\right).
\end{align*}
Hence a positive constant $\mathsf{c}$ independent of $M$ can be found such that $\VAR(\xi_n)\leq\mathsf{c}\frac{1}{M}\Rightarrow\VAR(\xi_n)= O\left(\frac{1}{M}\right)$.
\end{IEEEproof}
Since the pairwise error probabilities are $O(\frac{1}{M})$, from \eqref{eq:SERfloorFCNS} the SER floor at the high-SNR for the FC-NS case is also $O(\frac{1}{M})$. Hence, the SER floor for the FC-NS case can be made arbitrarily small as $M\rightarrow\infty$.

We conclude this section with a short intuitive explanation for the SER floors in both operations. At high-SNR in the S operation every BS antenna observes exactly the same signal. As a result, no advantage can be gained by using multiple antennas. In contrast, in the NS operation each BS antenna observes the symbol perturbed by some independent phase rotation. Hence, an averaging effect is observed for the independent phase noise sources at the BS array. Similar conclusions were also drawn in \cite{DurisiSLOCLO} for the uplink case.

We note that the single antenna transmitter is assumed to be phase noise free as our concern is the effect of the phase noise impairments at the BS array. The results are still valid in the case of a phase noise impaired transmitter for the S operation with appropriate adjustment of the notation. However, in the NS operation the averaging effect will not appear for the phase noise at the transmitter. Similar conclusions have already been drawn in prior work, such as \cite{TWireless}, where it is shown that in the NS operation the phase noise impairments at the BS average out but not the ones at the user terminals.

\section{Numerical Examples}\label{sec:NumericalExamples}

\begin{figure}[t!]
\centering
\begin{minipage}[t]{0.48\linewidth}
\psfrag{pi}{\scriptsize $\pi$}
\psfrag{-pi}{\scriptsize $-\pi$}
\psfrag{pi/2}{\scriptsize $\pi/2$} 
\psfrag{-pi/2}{\scriptsize $-\pi/2$} 
\includegraphics[width=\textwidth]{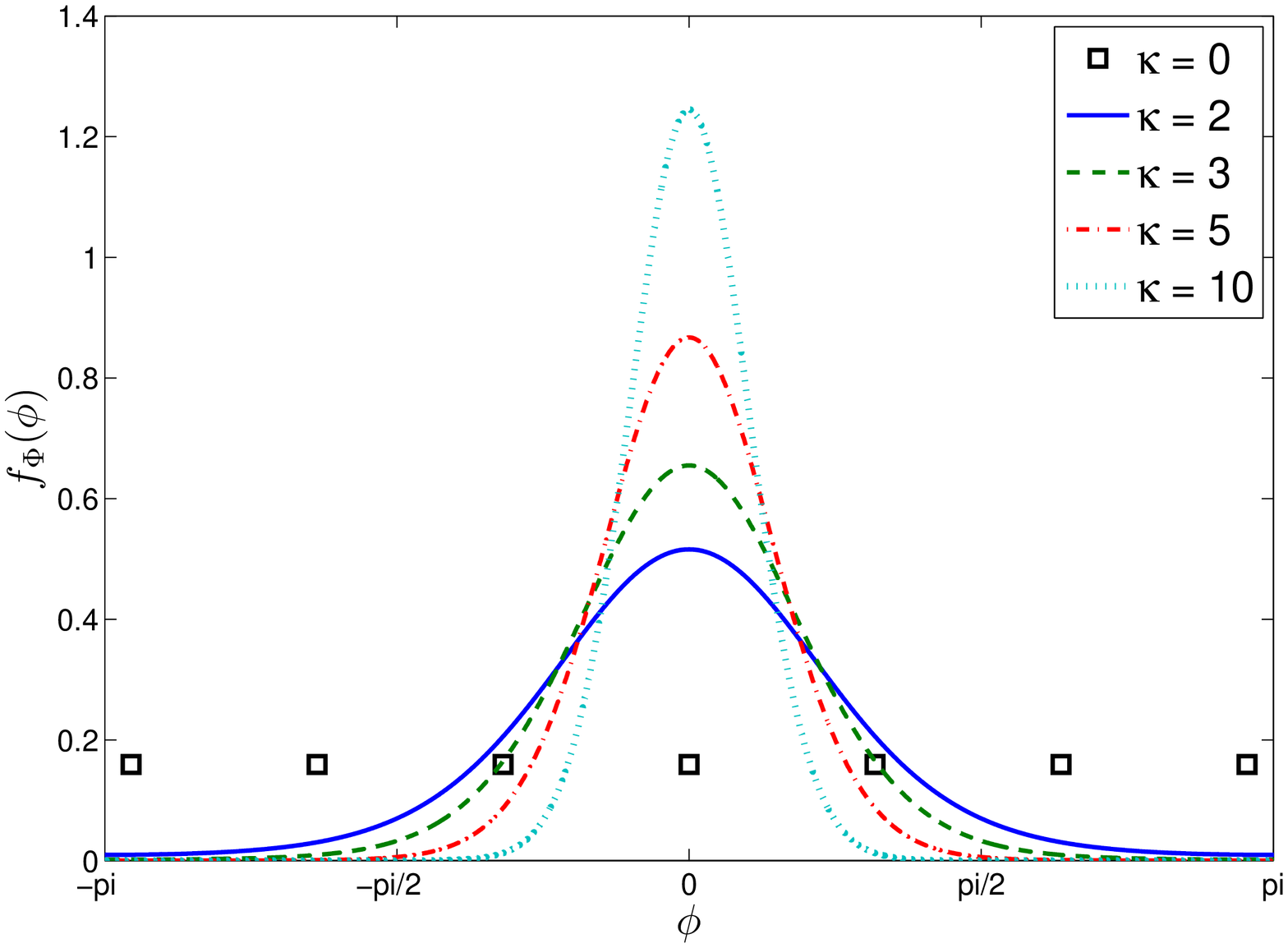}
\caption{The von Mises distribution, $\CircN(0,\kappa)$, for various choices of the parameter $\kappa$.}\label{fig:vonMises}
\end{minipage}
\quad
\begin{minipage}[t]{0.48\linewidth}
\includegraphics[width=\textwidth]{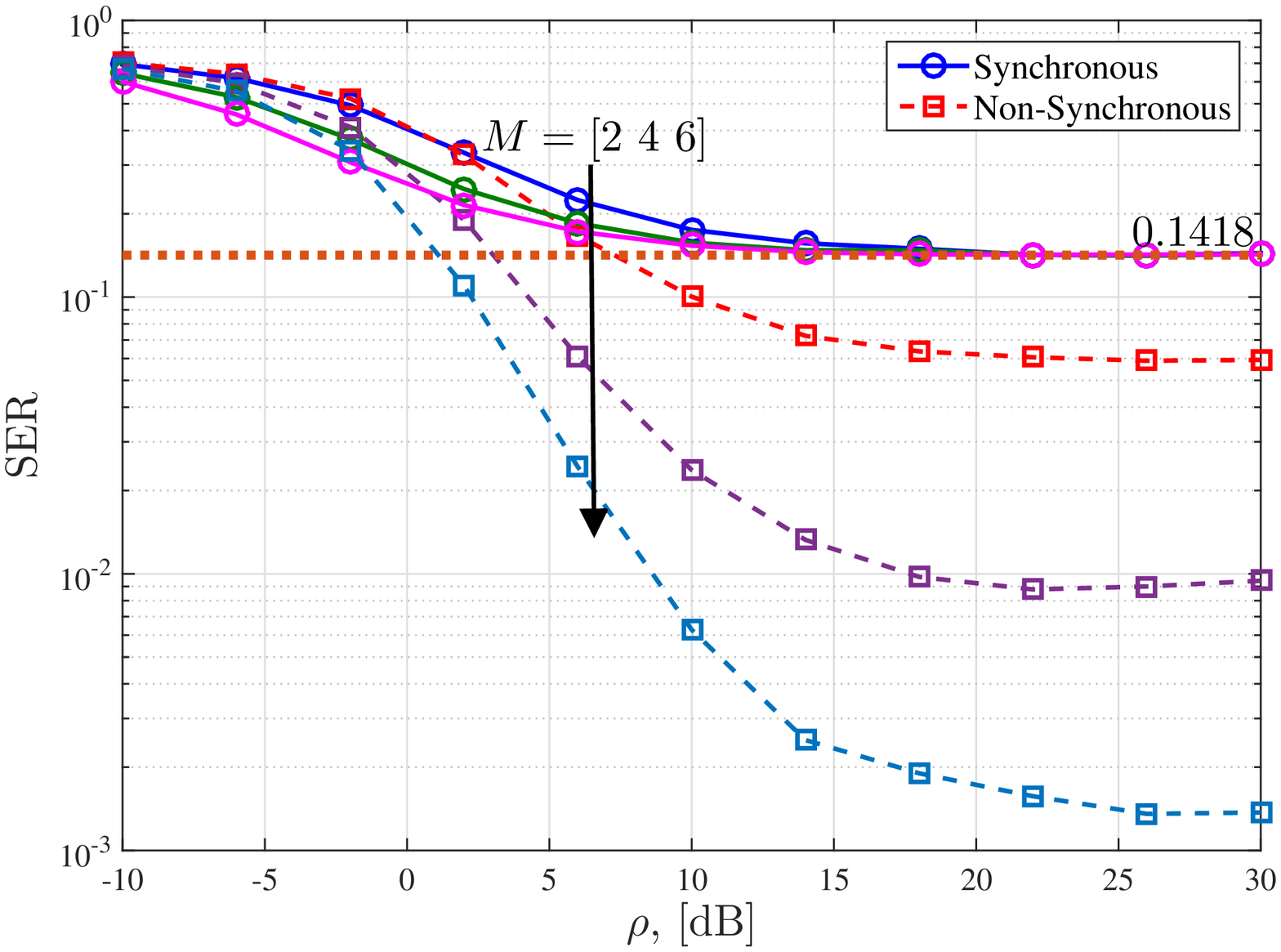}
\caption{SER performance as a function of the SNR $\rho$ for the constant channel case. The symbols are selected from a QPSK constellation and the concentration parameter is $\kappa=4$ for various values of $M$.}\label{fig:SERCCQPSKk4}
\end{minipage}
\end{figure}

In this section we present numerical examples that verify the validity of the analytical results presented in Sections \ref{sec:OptimalDetectors} and \ref{sec:HighSNR}. The examples are obtained by Monte Carlo simulations, where the receiver uses the detectors from Section \ref{sec:OptimalDetectors} to decide on the transmitted information symbols. In Fig. \ref{fig:vonMises} the zero mean von Mises distribution is plotted for reference purposes for various values of the concentration parameter $\kappa$. It is observed that the distribution is unimodal and that for $\kappa = 0$ it corresponds to the uniform distribution. As $\kappa$ increases the distribution becomes more concentrated around the mean. In the following we will refer to this distribution for our results, even though all the propositions and corollaries in Section \ref{sec:OptimalDetectors} hold for distributions that can be parameterized as in \eqref{eq:PhaseNoiseFourier}.

In Fig. \ref{fig:SERCCQPSKk4} the SER performance as a function of the SNR $\rho$ [dB] is plotted for $\kappa=4$ and $M\in\{2,4,6\}$ for the constant channel case. In all the numerical examples for the CC case is assumed to be $\bsg=[1,\ldots,1]^T$. In the low SNR regime array gains are observed for both operations and the synchronous operation is marginally better than the non-synchronous operation. However, the performance of both operations in that regime prohibits reliable communication. In the medium SNR regime ($\approx 0-10$ [dB]) the non-synchronous operation has a clear advantage over the synchronous operation. SER floors in the high-SNR regime are observed in all cases. Specifically for the synchronous operation the SER performance saturates at the same value irrespective of $M$. This is in line with Proposition \ref{prop:CCShighSNRErrorFloor}. The dotted straight line corresponds to the theoretically calculated SER floor, \eqref{eq:CCShighSNRErrorFloor}. The theoretical value $0.1418$ is also shown. Superior high-SNR performance is observed for the non-synchronous operation. In this case the SER floor is reduced when increasing the number of BS antennas. This verifies the results in Sections \ref{sub:HighSNRCCNS} and \ref{sub:highSNRFCNS}.

\begin{figure}[t!]
\centering
\begin{minipage}[t]{0.48\linewidth}
\includegraphics[width=\textwidth]{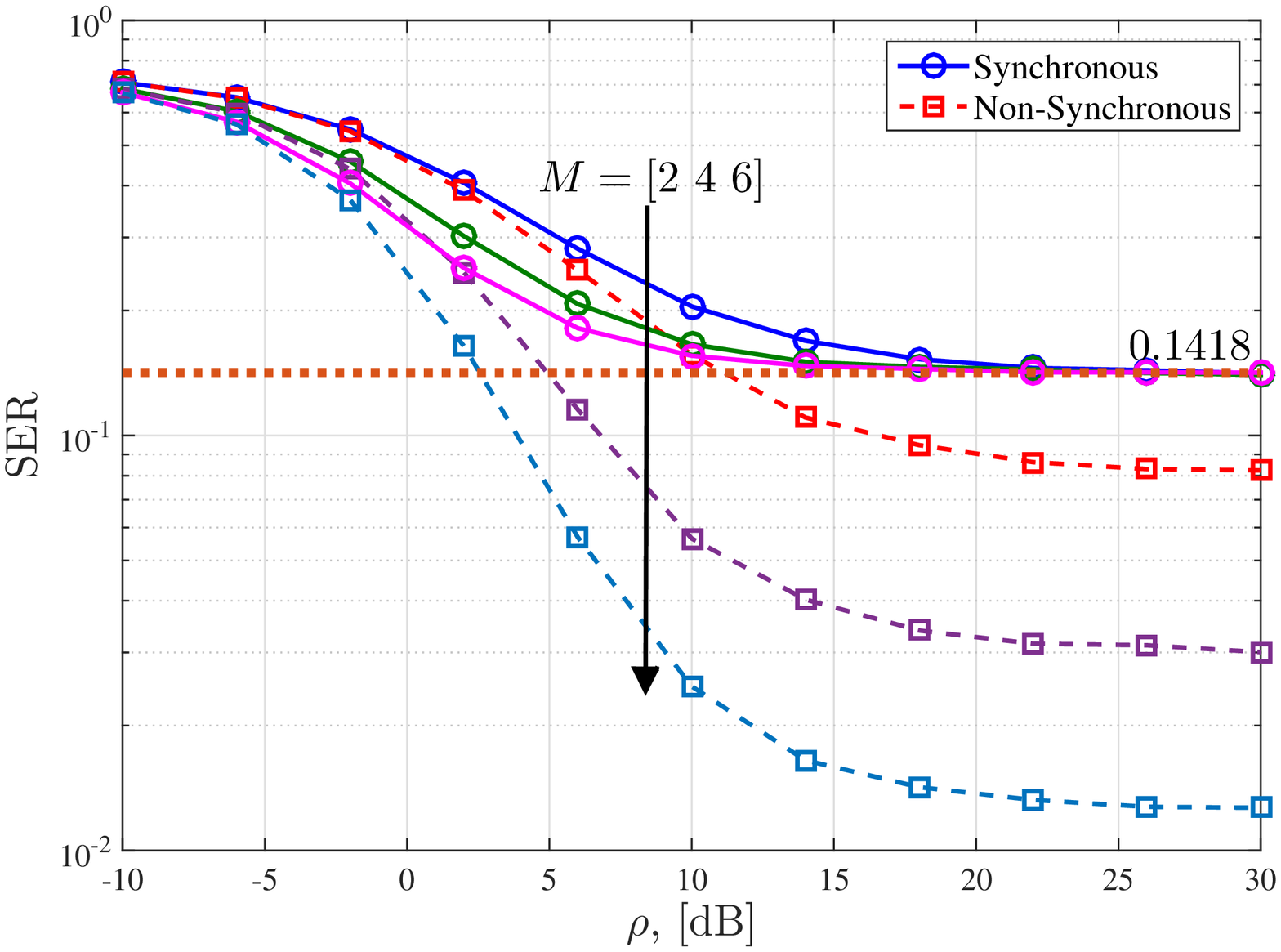}
\caption{SER performance as a function of the SNR $\rho$ for the fading channel case. The symbols are selected from a QPSK constellation and the concentration parameter is $\kappa=4$ for various values of $M$.}\label{fig:SERFCQPSKk4}
\end{minipage}
\quad
\begin{minipage}[t]{0.48\linewidth}
\includegraphics[width=\textwidth]{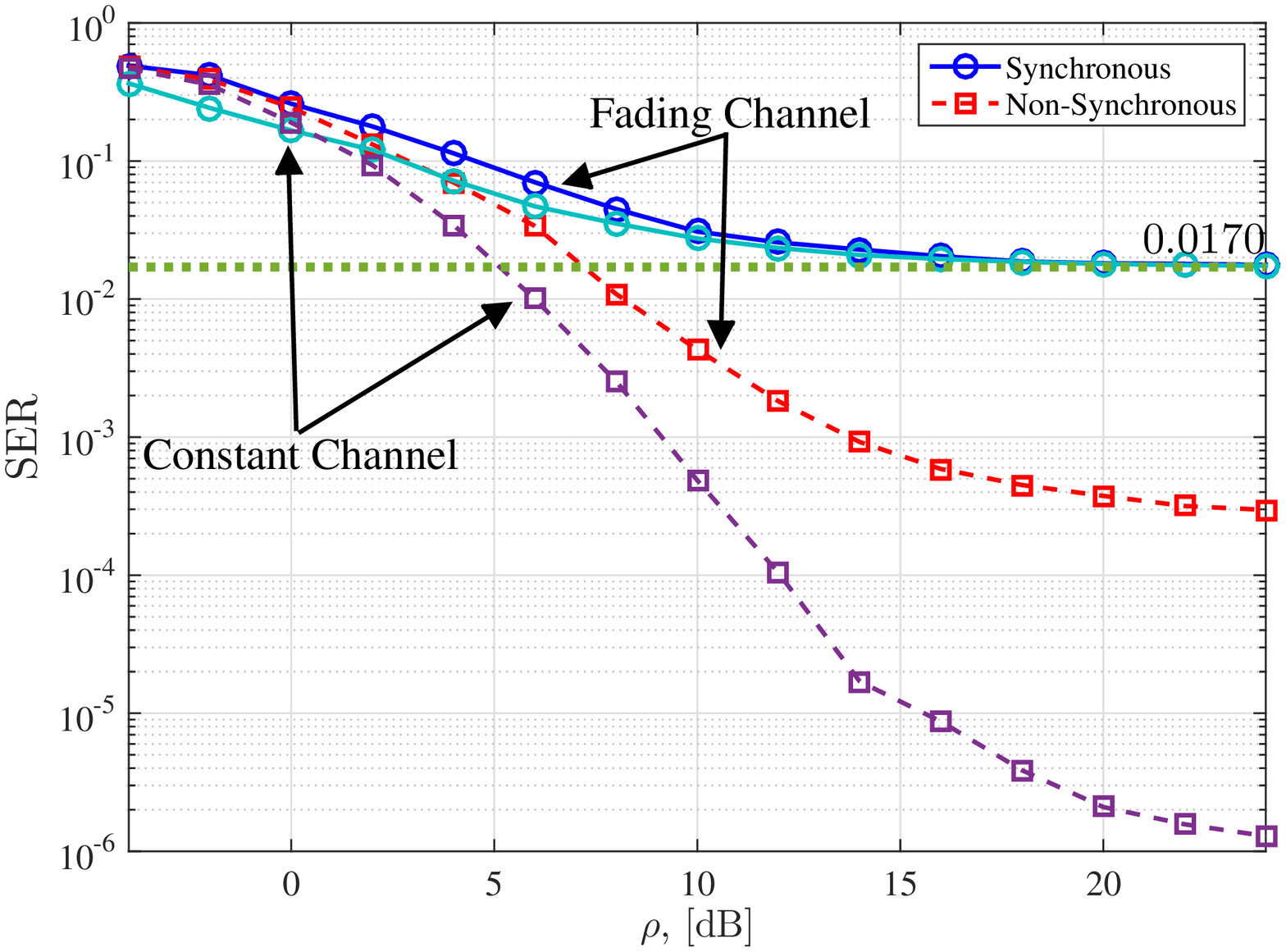}
\caption{Comparison of the SER performance as a function of the SNR $\rho$ for the CC and FC cases. The symbols are selected from a QPSK constellation and the concentration parameter is $\kappa=10$ and for $M=5$ BS antennas.}\label{fig:SERCC_FCQPSKM5k10}
\end{minipage}
\end{figure}

In Fig. \ref{fig:SERFCQPSKk4} the SER performance as a function of $\rho$ [dB] is plotted for $\kappa=4$ and $M\in\{2,4,6\}$ for the fading channel case. For the synchronous operation, an array gain is observed in the low SNR regime and the SER floor is the same for all the values of $M$. These observations are identical to the constant channel case and are in line with Proposition \ref{prop:CCShighSNRErrorFloor}. The theoretical SER floor is also plotted as in Fig. \ref{fig:SERCCQPSKk4}. Analogously to the constant channel case, SER floors are observed in the FC-NS case as well. Further, the SER floor is reduced by increasing the number of BS antennas. This is also in line with the results in Sections \ref{sub:HighSNRCCNS} and \ref{sub:highSNRFCNS}.
For the Figs. \ref{fig:SERCCQPSKk4} and \ref{fig:SERFCQPSKk4} the detectors in \eqref{eq:DetectionRuleNSgeneral} and \eqref{eq:DetectionRuleSgeneral} were computed so that the relative accuracy, $\delta_{\mathsf{acc}}(\nu)$, of the truncated metric with $\nu$ terms for the $s$ constellation symbol, $\tilde{\mathcal{L}}_s(\nu)$, defined as $\delta_{\mathsf{acc}}(\nu)\Define\left|\frac{\tilde{\mathcal{L}}_s(\nu)-\tilde{\mathcal{L}}_s(\nu-1)}{\tilde{\mathcal{L}}_s(\nu-1)}\right|$, was less than $10^{-12}$. In Table \ref{tab:NumberOfTerms} we summarize the mean and the maximum number of terms required to achieve the relative accuracy, $\delta_{\mathsf{acc}}(\nu)$, for a set of values of $\rho$ [dB]. We observe that as $\rho$ increases the number of terms required increases as well. However, in all cases less than 20 terms were sufficient for the desired relative accuracy. This demonstrates the fact that with only a few terms included in \eqref{eq:DetectionRuleNSgeneral} and \eqref{eq:DetectionRuleSgeneral}, the truncation error becomes negligible.
\begin{table}
\centering
\begin{tabular}{cc|c|c|c|c|c|c|}
\cline{3-8}
\multicolumn{1}{l}{}                        & \multicolumn{1}{l|}{} & \multicolumn{2}{c|}{$2$ dB} & \multicolumn{2}{c|}{$10$ dB} & \multicolumn{2}{c|}{$22$ dB} \\ \cline{3-8} 
\multicolumn{1}{l}{}                        & \multicolumn{1}{l|}{} & CC          & FC         & CC          & FC          & CC          & FC          \\ \hline
\multicolumn{1}{|c|}{\multirow{2}{*}{Mean}} & S                     & 13.8        & 12.7       & 16.2        & 16.2        & 15.7        & 17.1        \\ \cline{2-8} 
\multicolumn{1}{|c|}{}                      & NS                    & 10          & 12.5       & 13.8        & 16          & 15.8        & 16.6        \\ \hline
\multicolumn{1}{|c|}{\multirow{2}{*}{Max}}  & S                     & 17          & 16         & 17          & 17          & 16          & 18          \\ \cline{2-8} 
\multicolumn{1}{|c|}{}                      & NS                    & 13          & 16         & 17          & 17          & 17            & 18          \\ \hline
\end{tabular}
\caption{Mean and maximum number of terms, $\nu$, in the truncated sums of the optimal detectors from \eqref{eq:DetectionRuleNSgeneral} and \eqref{eq:DetectionRuleSgeneral} for relative accuracy of $\delta_{\mathsf{acc}}(\nu)<10^{-12}$, $M=6$ and $\kappa=4$.}\label{tab:NumberOfTerms}
\end{table}

In Fig. \ref{fig:SERCC_FCQPSKM5k10} the SER performance is plotted as a function of the SNR $\rho$ [dB] for the constant channel and fading channel cases. Similar performance is observed for the synchronous operation in both CC-S and FC-S cases. In the low SNR the constant channel exhibits better performance. This is due to the randomness of the fading channel case. However, as the number of BS antennas increases the squared norm of the fading channel vector normalized by $M$ becomes almost deterministic, so this gap is expected to be reduced. In the high-SNR regime the SER floor is the same for both the constant channel and the fading channel case. This is in agreement with Remark \ref{rem:EquivalenceCCFCSynch}. The theoretical SER floor for the CC-S and FC-S cases is also given by the dotted line. For the non-synchronous operation improved performance is observed in the constant channel case. This implies that the additional randomness due to fading in the FC-NS case has a direct impact on the SER performance. This is an expected behavior. In the absence of phase noise the CC-NS case corresponds to the detection in an AWGN channel with constant gain and the error probability can be expressed in terms of the $Q$-function. In the absence of phase noise, the FC-NS case corresponds to the coherent detection in a Rayleigh SIMO channel with perfect channel knowledge and the error probability scales only as $\rho^{-M}$\cite{TseBook}. This phenomenon naturally carries over when phase noise impairments are present.

\subsection{Extension to Longer Data Intervals}\label{subsec:Extension}

In practice more than one channel uses are spent for data transmission. In this section we demonstrate that the conclusions drawn by the study of the models in \eqref{eq:ConstantChannelMatrixVector} and \eqref{eq:FadingChannelMatrixVector} are valid for transmission protocols with multiple channel uses. For this purpose, we consider a setup where the data interval is extended to $\sfT$ channel uses. The calculation of the optimal detectors in closed form appears to be intractable \cite{Kam94TComm},\cite{Colavolpe05JSAC}, \cite{Rajet13MIMOphaseNoise}, \cite{Mehrpouyan12}. Hence, the approach can be summarized as follows. If a suboptimal tractable detector for the NS operation performs better than a genie-aided, i.e. \emph{better-than-optimal}, tractable detector for the S operation, then the same will hold for the corresponding optimal detectors.

We describe the approach for the FC case as the CC case follows in a similar fashion. For the FC-NS case the second equation of \eqref{eq:FadingChannelMatrixVector} is extended to $\bsy_t=\sqrt{\rho}\bsPhi_t\bsh s_t+\bsz_t$, for $t=1,\ldots,\sfT$, where the $m$-th element of the diagonal matrix $\bsPhi_t$ is $e^{j\sum_{\tau=1}^{t}\phi_m[\tau]}$ and the increments $\phi_m[\tau]$ are i.i.d. distributed according to \eqref{eq:PhaseNoiseFourier}. In the following we provide a suboptimal causal SBS detector with decision feedback for the symbol $s_t$. At time $t$ the information symbols up to $t-1$ have been detected and this information is used by the detector as if the detected values were the true ones. Further, for simplicity the detector substitutes the random rotations of the phase noise rotations at time $1\leq\tau\leq t-1$ with their statistical mean $d_m[\tau]=\Ebb\left[e^{j\sum_{\hat \tau=1}^{\tau}\phi_m[\hat \tau]}\right]=\alpha_{1,\tau}$, where we make the simplification that the phase noise statistics along the BS antennas are the same. Hence, the detector assumes that at some $\tau\in[1,t-1]$ the received vector $\bsy_\tau$ is observed via $\bsy_\tau=\sqrt{\rho}\diag\left\{d_1[\tau],\ldots,d_M[\tau]\right\}\bsh\hat s_\tau+\bsz_\tau$. Finally, at time $t$ the phase noise impairment is assumed to have the same statistics as the accumulated phase noise increments up to $t$, $\sum_{\tau=1}^{t}\phi_m[\tau]$. Hence the causal decision feedback detector selects the $s_t\in\mathcal{S}$ that maximizes the likelihood $p(\bsx,\bsy_1,\ldots,\bsy_t|\hat s_1,\ldots,\hat s_{t-1},s_t)$. The final result follows by similar steps as in Proposition \ref{prop:NSLikelihoodFunction}, i.e.
\begin{align}\label{eq:ExtendedFCDFDetector}
\hat s_t&=\argmax_{s\in\mathcal{S}}\hspace{5mm}-M\ln\left( \hat a_t+\rho|s_t|^2\right)+\frac{\rho\left(\sum_{m=1}^{M}|\hat v_m[t]|^2+|s_t|^2\|\bsy_t\|^2\right)}{\hat a_t+\rho|s_t|^2}\\
&+\sum_{m=1}^{M}\ln\bigg(\alpha_{0,t}I_0\left(\frac{2\rho|\hat\chi_m[t]|}{\hat a_t+\rho|s_t|^2}\right)+2\sum_{p=1}^{\infty}\alpha_{p,t}I_p\left(\frac{2\rho|\hat\chi_m[t]|}{\hat a_t+\rho|s_t|^2}\right)\cos\left(p\left(\arg\left(\hat\chi_m[t]\right)\right)\right)\bigg)\nonumber,
\end{align}
where $\hat a_t\Define1+\rho+\rho\sum_{\tau=1}^{t-1}|d_m[\tau]\hat s_\tau|^2$, $\hat v_m[t]\Define x_m^*+\sum_{\tau=1}^{t-1}y_m^*[\tau]d_m[\tau]\hat s_\tau$, $\hat\chi_m[t]\Define \hat v_{m,t}^*y_m^*[t]s_t$. The sequence $\alpha_{0,t},\alpha_{1,t},\ldots$ is the Fourier expansion of the pdf of $\sum_{\tau=1}^{t}\phi_m[\tau]$. As noted, this detector is suboptimal but implementable. Hence, the optimal causal detector for the FC-NS case must perform at least as good as this suboptimal detector.

\begin{figure}[t!]
\centering
\includegraphics[width=0.48\textwidth]{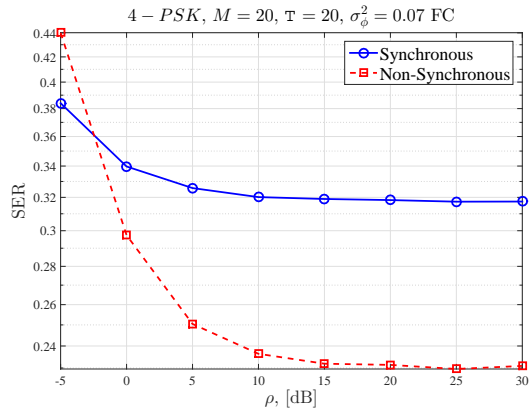}
\caption{Comparison of the detectors \eqref{eq:GenieAidedDetectorFCS} and \eqref{eq:ExtendedFCDFDetector} for $M=20$ and $\sfT=20$.}\label{fig:ExtendedSystemFC}
\end{figure}

In the following, we provide a genie-aided detector for the FC-S case. In this case, at time $t$ the received vector is given by $\bsy_t=\sqrt{\rho}e^{j\sum_{\tau=1}^{t}\phi[\tau]}\bsh s_t+\bsz_t$. Assume that at time $t$, the receiver is aware of $\vartheta[t]\Define\sum_{\tau=1}^{t-1}\phi[\tau]$ and the true $s_1,\ldots, s_{t-1}$. The causal ML detector selects the symbol $s_t\in\mathcal{S}$ that maximizes the likelihood $p(\bsx,\bsy_1,\ldots,\bsy_t|s_1,\ldots,s_t,\vartheta[t])$, i.e. 
\vspace{-5mm}\begin{align}\label{eq:GenieAidedDetectorFCS}
\hat s_t&=\argmax_{s_t\in\mathcal{S}}\hspace{5mm} -M\ln\left(a_t+\rho|s_t|^2\right)+\frac{\rho\left(\left\|\bsv_{t-1}\right\|^2+|s_t|^2\left\|\bsy_t\right\|^2\right)}{a_t+\rho|s_t|^2}\\
&+\ln\left(\alpha_0I_0\left(\frac{2\rho|\chi_t|}{a_t+\rho|s_t|^2}\right)+2\sum_{p=1}^{\infty}\alpha_pI_p\left(\frac{2\rho|\chi_t|}{a_t+\rho|s_t|^2}\right)\cos\left(p\left(\vartheta[t]-\arg\left(\chi_t\right)\right)\right)\right)\nonumber,
\end{align}
where $a_t\Define 1+\rho+\rho\sum_{\tau=1}^{t-1}|s_\tau|^2$, $\bsv_{t-1}\Define\bsx+\sum_{\tau=1}^{t-1}s_\tau^*e^{-j(\theta_\tau+\phi_\tau)}\bsy_\tau$ and $\chi_t\Define s_t^*\bsv_{t-1}^H\bsy_t.$ This detector performs better than the actual causal ML detector for the synchronous operation, since it has the additional knowledge of the true prior symbols and the evolution of the phase noise process up to $t-1$. In Fig. \ref{fig:ExtendedSystemFC} the SER of the two detectors in \eqref{eq:GenieAidedDetectorFCS} and \eqref{eq:ExtendedFCDFDetector} is plotted for data interval length $\sfT=20$ and QPSK symbols. The phase noise increments $\phi_m[t]$ are assumed to be i.i.d. zero mean wrapped Gaussian with variance $\sigma_\phi^2=0.07$. This corresponds to the practical scenario of free-running oscillators. In this case, the pdf given by \eqref{eq:PhaseNoiseFourier}, where $\alpha_{m,p}=\left(\exp\left(-\frac{\sigma_\phi^2}{2}\right)\right)^{p^2}$ \cite{CircularStatisticsBook}. It is clear that the suboptimal FC-NS detector performs better than the 'better-than-optimal' FC-S detector, which establishes the claim that the results of the previous sections are valid in more complex setups.

\section{Conclusions}\label{sec:Conclusions}
The problem of ML detection in a training-assisted single-user SIMO channel with phase noise impairments and $M$ BS antennas was studied. A simple transmission protocol was considered, where one time slot is used for pilot and the second time slot is used for data transmission. The study included the case of constant and fading channel assumptions. For both assumptions on the channel gains two operations were investigated, i.e., the synchronous and non-synchronous operations. Closed-form expressions of the optimal detectors were given for a general parameterization of the phase noise increments. SER floors were observed for all cases under study. For the synchronous operation the SER floors were independent of $M$ for both the constant and fading channel case. In contrast the SER floor in the non-synchronous operation could be made arbitrarily small by increasing $M$. 

The effects discovered in the paper are fundamental because they assume optimal detection in the Bayesian sense, as the detector was found by analytical marginalization of all nuisance parameters, i.e., the phase noise in the CC case and the channel fading and the phase noise in the FC case. That is, no filtering algorithm, alternative or additional, can improve the SER performance of any of the operations, S and NS. The observed SER floors and the distinction between S and NS operation are, thus, fundamental and not an artifact of suboptimal receiver processing. Further, we have shown that our results remain valid if the data interval is extended to multiple channel uses.

\appendices

\vspace{-8mm}
\section{Two-Slot Proofs}

\subsection{Proof of Proposition \ref{prop:NSLikelihoodFunction} for the CC-NS case}\label{app:CCNSLikelihoodFunction}

Define the vectors $\bstheta\Define[\theta_1,\ldots,\theta_M]^T$ and $\bsphi\Define[\phi_1,\ldots,\phi_M]^T$. The likelihood function is given by
\begin{align}\label{eq:CCNSLikelihoodFactorization}
p(\bsx,\bsy|s)&=\iint p(\bsx,\bsy|s,\bstheta,\bsphi)p(\bstheta,\bsphi|s)\diff  \bstheta \diff \bsphi\stackrel{(a)}{=}\iint p(\bsx|\bstheta)p(\bsy|s,\bstheta,\bsphi)p(\bstheta)p(\bsphi)\diff \bstheta \diff \bsphi\nonumber\\
&\stackrel{(b)}{=}\prod_{m=1}^{M}\int_{-\pi}^{\pi}p(x_m|\theta_m)p(\theta_m)\underbrace{\int_{-\pi}^{\pi}p(y_m|s,\theta_m,\phi_m)p(\phi_m)\diff \phi_m}_{\Define p(y_m|\theta_m,s)} \diff \theta_m=\prod_{n=1}^{M}p(x_m,y_m|s),
\end{align}
where (a) follows from the fact that conditioned on $\bstheta,~\bsphi$ and $s$, the vectors $\bsx$ and $\bsy$ are independent and that $\bstheta,~\bsphi$ are independent of each other and $s$. The equality in (b) is a consequence of the independence of the components in $\bstheta$, $\bsphi$, $\bsw$ and $\bsz$. The channel probability law $p(y_m|s,\theta_m,\phi_m)$ can be expressed as 
\begin{align}\label{eq:InnerChannelLaw}
p(y_m&|s,\theta_m,\phi_m)=\frac{e^{-\left|y_m-\sqrt{\rho}g_me^{j(\theta_m+\phi_m)}s\right|^2}}{\pi}=\frac{e^{-|y_m|^2-\rho g_m^2|s|^2+2\sqrt{\rho}g_m|s^*y_m|\cos\left(\phi_m+\theta_m-\arg\left(s^*y_m\right)\right)}}{\pi}\nonumber\\
&=\frac{e^{-|y_m|^2-\rho g_m^2 |s|^2}}{\pi}\left(I_0(2\sqrt{\rho}g_m|s^*y_m|)+2\sum_{l=1}^{\infty}I_l(2\sqrt{\rho}g_m|s^*y_m|)\cos\left(l\left(\phi_m+\theta_m-\arg\left(s^*y_m\right)\right)\right)\right),
\end{align}
where the last step follows from the Jacobi-Anger formula \cite[Section 4.4, p. 100]{viterbi66coherentCommunication}
\begin{align}\label{eq:JacobiAnger}
e^{\alpha\cos\beta}=I_0(\alpha)+2\sum_{l=1}^{\infty}I_l(\alpha)\cos(l\beta).
\end{align}
Then, the conditional $p(y_m|\theta_m,s)$ is
\begin{align*}
p&(y_m|\theta_m,s)=\int_{-\pi}^{\pi}p(y_m|s,\theta_m,\phi_m)p(\phi_m)\diff \phi_m\\
&=\frac{e^{-|y_m|^2-\rho g_m^2 |s|^2}}{\pi}\left(\alpha_{m,0} I_0(2\sqrt{\rho}g_m|s^*y_m|)+2\sum_{l=1}^{\infty}\alpha_{m,l} I_l(2\sqrt{\rho}g_m|s^*y_m|)\cos\left(l\left(\theta_m-\arg\left(s^*y_m\right)\right)\right)\right).
\end{align*}
By manipulating $p(x_m|\theta_m)$ in the same way as in \eqref{eq:InnerChannelLaw} and by the orthogonality of $\cos$ and $\sin$ we obtain
\begin{align}\label{eq:partialLikelihoodsNonSynch}
p(x_m,y_m|s)&=\int_{-\pi}^{\pi}p(x_m|\theta_m)p(\theta_m)p(y_m|\theta_m,s) \diff \theta_m\nonumber\\
&=\frac{e^{-|x_m|^2-|y_m|^2-\rho g_m^2 (1+|s|^2)}}{\pi^2}\left(\beta_{m,0}+2\sum_{l=1}^{\infty}\beta_{m,l}\cos\left(l\zeta_m\right)\right).
\end{align}
The result in \eqref{eq:NSLikelihoodFunction}, \eqref{eq:ADefinitionCC}, \eqref{eq:betaDefinitionCCNS} and \eqref{eq:zetaDefinitionNS} follows by substituting \eqref{eq:partialLikelihoodsNonSynch} in \eqref{eq:CCNSLikelihoodFactorization}.

\subsection{Proof of Proposition \ref{prop:NSLikelihoodFunction} for the FC-NS case}\label{app:FCNSLikelihoodFunction}

The likelihood function in this case is given by
\begin{align}\label{eq:FCNSLikelihoodFactorization}
p(\bsx,&\bsy|s)=\int p(\bsx,\bsy|s,\bsphi,\bsh)p(\bsh,\bsphi|s)\diff \bsh \diff \bsphi\stackrel{(a)}{=}\int p(\bsx|\bsh)p(\bsy|s,\bsphi,\bsh)p(\bsh)p(\bsphi) \diff\bsh \diff \bsphi\nonumber\\
&=\int p(\bsphi)\int p(\bsh)p(\bsy|s,\bsphi,\bsh)p(\bsx|\bsh)\diff \bsh\diff \bsphi\nonumber\\
&\stackrel{(b)}{=}\prod_{m=1}^{M}\int_{-\pi}^{\pi}p(\phi_m) \underbrace{\int_{\mathbb{C}} p(h_m)p(y_m|s,\phi_m,h_m)p(x_m|h_m)\diff h_m}_{\Define p(x_m,y_m|s,\phi_m)}\diff \phi_m,
\end{align}
where $(a)$ follows from the fact that conditioned on $s,~\bsh$ and $\bsphi$, the received vectors $\bsx$ and $\bsy$ are independent and from the assumption that $s,~\bsh$ and $\bsphi$ are mutually independent. Further, the factorization in $(b)$ follows from the fact that the components of $\bsh$ and $\bsphi$ are independent. It holds that
\begin{align}\label{eq:FCNSTrainingPDF}
p(x_m|h_m)&=\frac{e^{-\left|x_m-\sqrt{\rho}h_m\right|^2}}{\pi}=\frac{e^{-|x_m|^2-\rho|h_m|^2}}{\pi}\exp\left(\Re\left\{2\sqrt{\rho}x_m^*h_m \right\}\right),
\end{align}
\begin{align}\label{eq:FCNSDataPDF}
p(y_m|\phi_m,h_m,s)&=\frac{e^{-|y_m-\sqrt{\rho}e^{j\phi_m}h_m s|^2}}{\pi}=\frac{e^{-|y_m|^2-\rho|s|^2|h_m|^2}}{\pi}\exp\left(\Re\left\{2\sqrt{\rho}y_m^*h_m s e^{j\phi_m}\right\}\right).
\end{align}
From \eqref{eq:FCNSTrainingPDF}, \eqref{eq:FCNSDataPDF} and $p(h_m)=\frac{1}{\pi}e^{-|h_m|^2}$ the conditional pdf $p(x_m,y_m|s,\phi_m)$ is written as
\begin{align*}
&p(x_m,y_m|s,\phi_m)=\int_{\mathbb{C}}p(h_m)p(y_m|s,\phi_m,h_m)p(x_m|h_m)\diff h_m\\
&=\frac{e^{-|x_m|^2-|y_m|^2}}{\pi^{3}}\int_{\mathbb{C}}\!\exp\!\left(-\left(\rho+\rho|s|^2+1\right)\left[|h_m|^2-2\Re\left\{\frac{\sqrt{\rho}\left(x_m+s^*y_me^{-j\phi_m}\right)^*h_m}{\rho+\rho|s|^2+1}\right\}\right]\right)\diff h_m
\end{align*}
For notational convenience we define $c_m\Define\frac{\sqrt{\rho}\left(x_m+s^*y_me^{-j\phi_m}\right)}{\rho+\rho|s|^2+1}$ and $p(x_m,y_m|s,\phi_m)$ is expressed as
\begin{align*}
&p(x_m,y_m|s,\phi_m)=\frac{e^{-|x_m|^2-|y_m|^2}}{\pi^{3}}\int_{\mathbb{C}}e^{-\left(\rho+\rho|s|^2+1\right)\left[|h_m|^2-2\Re\left\{c_m^*h_m\right\}+|c_m|^2-|c_m|^2\right]}\diff h_m\\
&\stackrel{(a)}{=}\frac{e^{-|x_m|^2-|y_m|^2+\left(\rho+\rho|s|^2+1\right)|c_m|^2}}{\pi^{3}}\int_{\mathbb{C}}e^{-\left(\rho+\rho|s|^2+1\right)\left|h_m-c_m\right|^2}\diff h_m\stackrel{(b)}{=}\frac{e^{-|x_m|^2-|y_m|^2+\left(\rho+\rho|s|^2+1\right)|c_m|^2}}{\pi^2\left(\rho+\rho|s|^2+1\right)},
\end{align*}
where in $(a)$ we complete the square and $(b)$ is the integral of a complex Gaussian function over $\mathbb{C}$. By using these expressions we can compute
\begin{align}\label{eq:FCNSpdfIntermediate}
& p(x_m,y_m|s)=\int_{-\pi}^{\pi}p(\phi_m)p(x_m,y_m|s,\phi_m)\diff\phi_m\nonumber\\
&=\frac{e^{-\frac{1+\rho|s|^2}{\rho+\rho|s|^2+1}|x_m|^2-\frac{1+\rho}{\rho+\rho|s|^2+1}|y_m|^2}}{\pi^2\left(\rho+\rho|s|^2+1\right)}\int_{-\pi}^{\pi}p(\phi_m)e^{\frac{2\rho|s^*x_m^*y_m|}{\rho+\rho|s|^2+1}\cos\left(\arg(x_m^*y_m)-\phi_m-\arg(s)\right)}\diff\phi_m.
\end{align}
For convenience of the notation, we define
\begin{align}\label{eq:FCNSIntermediateDefinitions}
A_{m,1}\Define\frac{e^{-\frac{1+\rho|s|^2}{\rho+\rho|s|^2+1}|x_m|^2-\frac{1+\rho}{\rho+\rho|s|^2+1}|y_m|^2}}{\pi^2\left(\rho+\rho|s|^2+1\right)},~A_{m,2}\Define\frac{2\rho|s^*x_m^*y_m|}{\rho+\rho|s|^2+1},\text{ and }A_{m,3}\Define\arg(s)-\arg(x_m^*y_m).
\end{align}
By substituting \eqref{eq:PhaseNoiseFourier} and \eqref{eq:FCNSIntermediateDefinitions} into \eqref{eq:FCNSpdfIntermediate}
\begin{align}\label{eq:FCNSfinalStep}
& p(x_m,y_m|s)=\frac{A_{m,1}}{2\pi}\int_{-\pi}^{\pi}\left(\alpha_{m,0}+2\sum_{l=1}^{\infty}\alpha_{m,l}\cos\left(l\phi_m\right)\right)e^{A_{m,2}\cos\left(\phi_m+A_{m,3}\right)}\diff\phi_m\nonumber\\
&=\frac{A_{m,1}\alpha_{m,0}}{2\pi}\int_{-\pi}^{\pi}e^{A_{m,2}\cos\left(\phi_m+A_{m,3}\right)}\diff\phi_m+\frac{A_{m,1}}{2\pi}\sum_{l=1}^\infty\int_{-\pi}^{\pi}2\alpha_{m,l}\cos\left(l\phi_m\right)e^{A_{m,2}\cos\left(\phi_m+A_{m,3}\right)}\diff\phi_m\nonumber\\
&=A_{m,1}\alpha_{m,0} I_0\left(A_{m,2}\right)\nonumber\\
&+\frac{A_{m,1}}{\pi}\sum_{l=1}^\infty\alpha_{m,l}\int_{-\pi}^{\pi}\cos\left(l\phi_m\right)\left(I_0(A_{m,2})+2\sum_{k=1}^{\infty}I_k(A_{m,2})\cos\left(k\left(\phi_m+A_{m,3}\right)\right)\right)\diff\phi_m\nonumber\\
&=A_{m,1}\left(\alpha_{m,0} I_0\left(A_{m,2}\right)+2\sum_{l=1}^\infty\alpha_{m,l} I_l(A_{m,2})\cos\left(lA_{m,3}\right)\right).
\end{align}
Since $p(x_m,y_m|s)$ is a density function and hence it is absolutely integrable, the interchange of summation and integration is possible by the Fubini-Tonelli theorem \cite{Billingsley}. The result in \eqref{eq:NSLikelihoodFunction}, \eqref{eq:ADefinitionFC}, \eqref{eq:betaDefinitionFCNS} and \eqref{eq:zetaDefinitionNS} follows by substituting \eqref{eq:FCNSIntermediateDefinitions} and \eqref{eq:FCNSfinalStep} into \eqref{eq:FCNSLikelihoodFactorization}.

\subsection{Proof of Proposition \ref{prop:SLikelihoodFunction} for the CC-S case}\label{app:CCSLikelihoodFunction}

The likelihood function in this case is factorized similarly to \eqref{eq:CCNSLikelihoodFactorization} as
\begin{align}\label{eq:CCSLikelihoodFactorization}
&p(\bsx,\bsy|s)=\int p(\bsx|\theta)p(\theta)\int p(\bsy|s,\theta,\phi)p(\phi)\diff \phi\diff \theta.
\end{align}
The integrals in \eqref{eq:CCSLikelihoodFactorization} can be calculated using a similar approach as in Appendix \ref{app:CCNSLikelihoodFunction} yielding \eqref{eq:SLikelihoodFunction}, \eqref{eq:ADefinitionCC}, \eqref{eq:betaDefinitionCCS} and \eqref{eq:zetaDefinitionCCS}.

\subsection{Proof of Proposition \ref{prop:SLikelihoodFunction} for the FC-S case}\label{app:FCSLikelihoodFunction}

The likelihood function in this case is factorized similarly to \eqref{eq:FCNSLikelihoodFactorization} as
\begin{align}\label{eq:FCSLikelihoodFactorization}
p(\bsx,\bsy|s)=\int_{-\pi}^{\pi}p(\phi)\int_{\mathbb{C}^M} p(\bsy|s,\phi,\bsh)p(\bsh) p(\bsx|\bsh)\diff \bsh\diff \phi.
\end{align}
The integrals in \eqref{eq:FCSLikelihoodFactorization} can be calculated using a similar approach as in Appendix \ref{app:FCNSLikelihoodFunction} yielding \eqref{eq:SLikelihoodFunction}, \eqref{eq:ADefinitionFC}, \eqref{eq:betaDefinitionFCS} and \eqref{eq:zetaDefinitionFCS}.

\subsection{Proof of Proposition \ref{prop:vonMisesFCDetector}}\label{app:vonMisesFCDetector}

The proof is similar to the proof in Appendix \ref{app:FCNSLikelihoodFunction}. We substitute the von Mises pdf $p(\phi_m)=\frac{e^{\kappa\cos\phi_m}}{2\pi I_0(\kappa)}$ into \eqref{eq:FCNSpdfIntermediate} and evaluate the resulting integral. This yields \eqref{eq:vonMisesFCNSDetector}. The proof for \eqref{eq:vonMisesFCSDetector} is similar to the proof for \eqref{eq:vonMisesFCNSDetector}.

\section{$\sfT$-slot Detectors}

In this appendix detectors are presented for the $\sfT$-data-slot extended transmission protocol. Due to the fact that the derivation of the optimal detectors in the maximum-likelihood (ML) sense in closed form appears to be intractable, suboptimal detectors for the non-synchronous (NS) operation and genie-aided, i.e. \emph{better-than-optimal}, detectors for the synchronous (S) operation are derived. By simulations it is shown that the performance of the suboptimal NS detectors is superior to the genie-aided S detectors. This conclusively establishes the claim that the results of the two-slot model are valid for the extended models as well.


\subsection{Decision Feedback Detector: Non-Synchronous Operation}\label{subsec:NonSynchronousCC}

We consider a transmission interval of $\sfT+1$ channel uses. Training is done during the first channel use and uncoded information symbols are transmitted during the subsequent $\sfT$ channel uses. The received vector during training, $\bsx$, and the received vectors during data transmission, $\bsy_t$, at time $t$ are given by
\begin{align}
\bsx&=\sqrt{\rho}\bsTheta\bsg + \bsw,\label{eq:ReceivedTrainingCCNS}\\
\bsy_t&=\sqrt{\rho}\bsTheta\bsPhi_t\bsg s_t + \bsz_t,\hspace{5mm} t = 1,\ldots,\sfT,\label{eq:ReceivedDataCCNS}
\end{align}
where $\bsTheta\Define\diag\left\{e^{j\theta_1},\ldots,e^{j\theta_M}\right\}$ and $\bsPhi_t\Define\diag\left\{e^{j\sum_{\tau=1}^{t}\phi_1[\tau]},\ldots,e^{j\sum_{\tau=1}^{t}\phi_M[\tau]}\right\}$. The phases $\theta_m,~m=1,\ldots,M$ are random i.i.d. initial phases uniformly distributed in $[-\pi,\pi)$. The phase noise increments $\phi_m[t]$ are i.i.d random variables with pdf parameterized by
\begin{align}\label{eq:PhaseNoiseIncrementsPDF}
p_{\Phi_m[\tau]}(\phi_m[\tau])=\frac{1}{2\pi}\left(\alpha_{m,0}+2\sum_{p=1}^{\infty}\alpha_{m,p}\cos\left(p\phi_m[\tau]\right)\right).
\end{align}
The vector $\bsg\in\mathbb{R}^M$ is a constant and known vector of amplitudes and $s_t$ are data symbols selected from a fixed constellation $\mathcal{S}$. In order to derive a suboptimal detector for the symbol $s_t$ we consider that the receiver uses the detected symbols $\hat s_1,\ldots,\hat s_{t-1}$ as if they were true. Further, the exact knowledge of the phase noise increments is not available and can be only estimated. For simplicity of the detector we assume that distortion due to phase noise is equal to the expected value of it conditioned on the matrix of the initial phases $\bsTheta$. Hence, the received vectors $\bsy_\tau,~1\leq\tau\leq t-1$ are assumed to be observed by
\begin{align}\label{eq:HeuristicChannelCCNS}
\bsy_\tau&=\sqrt{\rho}\bsTheta\bsD_\tau\bsg\hat s_\tau+\bsz_\tau,\hspace{5mm}1\leq\tau\leq t-1,
\end{align}
where $\bsD_\tau\Define\Ebb\left[\bsPhi_\tau\right]\Define\diag\left\{\alpha_{1,1,\tau},\ldots,\alpha_{M,1,\tau}\right\}$, where $\alpha_{m,1,\tau}$ are the coefficients of the Fourier expansion of the pdf of $\sum_{\hat \tau=1}^{\tau}\phi_m[\hat \tau]$. Finally, at time $t$ the received vector is given by \eqref{eq:ReceivedDataCCNS}, where the phase noise perturbation equals to the total accumulated phase noise up to time $t$, $\varphi_m[t]\Define\sum_{\tau=1}^t\phi_m[\tau]$. The suboptimal causal decision feedback receiver is given by
\begin{align}\label{eq:DFReceiverCCNS}
\hat s_t&=\argmax_{s_t\in\mathcal{S}} p(\bsx,\bsy_1^t|\hat s_1^{t-1},s_t),
\end{align}
where we denote the sequence $v_a^b\Define\{v_a,\ldots,v_b\}$. The detailed steps of the derivation follow. The likelihood $p(\bsx,\bsy_1^t|\hat s_1^{t-1},s_t)$ is written as
\begin{align*}
p(\bsx,\bsy_1^t|\hat s_1^{t-1},s_t)&=\idotsint p(\bsx,\bsy_1^t|\hat s_1^{t-1},s_t,\bsTheta,\bsPhi_t)p(\bsTheta)p(\bsPhi_t)\diff\bsTheta\diff\bsPhi_t\\
&=\prod_{m=1}^M\iint p(x_m,\{y_m[\tau]\}_1^t|\hat s_1^{t-1},s_t,\theta_{m},\varphi_{m}[t])p(\theta_{m})p(\varphi_{m}[t])\diff\theta_{m}\diff\varphi_{m}[t].
\end{align*}
The derivation proceeds with marginalizing over the accumulated phase noise $\varphi_m[t]$.
\begin{align*}
p(x_m,\{y_m[\tau]\}_1^t|\hat s_1^{t-1},s_t,\theta_{m})&=\int_{-\pi}^\pi p(x_m,\{y_m[\tau]\}_1^t|\hat s_1^{t-1},s_t,\theta_{m},\varphi_{m}[t])p(\varphi_{m}[t])\diff\varphi_{m}[t],
\end{align*}
where
\begin{align*}
p(\varphi_{m}[t])=\frac{1}{2\pi}\left(\alpha_{0,m,t}+2\sum_{p=1}^{\infty}\alpha_{p,m,t}\cos\left(p\varphi_{m}[t]\right)\right).
\end{align*}
Note that here the coefficients of the Fourier expansion of $p(\varphi_m[t])$ depend on $t$, since the larger the $t$ the larger the variance of $\varphi_m[t]$. The density $p(x_m,\{y_m[\tau]\}_1^t|\hat s_1^{t-1},s_t,\theta_{m},\varphi_{m}[t])$ is given by
\begin{align*}
p(x_m,\{y_m[\tau]\}_1^t|\hat s_1^{t-1},s_t,\theta_{m},\varphi_{m}[t])&=\frac{1}{\pi^{t+1}}e^{-|x_{m}-\sqrt{\rho}e^{j\theta_{m}}g_m|^2-\sum_{\tau=1}^{t-1}|y_m[\tau]-\sqrt{\rho}e^{j\theta_{m}}d_m[\tau]g_m\hat s_\tau|^2}\\
&\times e^{-|y_m[t]-\sqrt{\rho}e^{j\theta_{m}}e^{j\varphi_{m}[t]}g_m s_t|^2}.
\end{align*}
With the use of the Jacobi-Anger formula, the phase noise term $\varphi_m[t]$ can be marginalized out and the density $p(x_m,\{y_m[\tau]\}_1^t|\hat s_1^{t-1},s_t,\theta_{m})$ is given by
\begin{align*}
p&(x_m,\{y_m\}_1^t|\hat s_1^{t-1},s_t,\theta_{m})=\frac{1}{\pi^{t+1}}e^{-|x_m|^2-\sum_{\tau=1}^{t}|y_m[\tau]|^2-\rho g_m^2\left(b_m[t]+|s_t|^2\right)+2\sqrt{\rho}\left|c_m[t]g_m\right|\cos\left(\theta_m+\arg\left(c_m[t] g_m\right)\right)}\\
&\times\left(\alpha_{0,m,t}I_0(2\sqrt{\rho}|B_m[t]|)+2\sum_{q=1}^{\infty}\alpha_{q,m,t}I_q(2\sqrt{\rho}|B_m[t]|)\cos\left(q\left(\theta_m+\arg\left(B_m[t]\right)\right)\right)\right),
\end{align*}
where $B_m[t]\Define y_m^*[t]g_m s_t$, $b_m[t]\Define 1+\sum_{\tau=1}^{t-1}|d_m[\tau]\hat s_\tau|^2$, $c_m[t]\Define x_m^*+\sum_{\tau=1}^{t-1}y_m^*[\tau]d_m[\tau]\hat s_\tau$. Further marginalization with respect to $\theta_m$ can be carried out which yields the density
\begin{align*}
&p(x_m,\{y_m[\tau]\}_1^t|\hat s_1^{t-1},s_t)=\int_{-\pi}^\pi p(x_m,\{y_m[\tau]\}_0^t|\hat s_1^{t-1},s_t,\theta_m)p(\theta_m)\diff\theta_m\\
&=\frac{1}{\pi^{t+1}}e^{-|x_m|^2-\sum_{\tau=1}^{t}|y_m[\tau]|^2-\rho g_m^2\left(b_m[t]+|s_t|^2\right)}\bigg(\alpha_{0,m,t}I_0(2\sqrt{\rho}|B_m[t]|)I_0(2\sqrt{\rho}|c_m[t] g_m|)\\
&+2\sum_{p=1}^{\infty}\alpha_{p,m,t}I_p(2\sqrt{\rho}|B_m[t]|)I_p(2\sqrt{\rho}|c_m[t] g_m|)\cos\left(p\left(\arg\left(B_m[t]\right)-\arg\left(c_m[t] g_m\right)\right)\right)\bigg).
\end{align*}
The detector is finally given by
\begin{align}\label{eq:DecisionFeedbackCCNS}
\hat s_t&=\argmax_{s_t\in\mathcal{S}}\hspace{5mm} -\rho|s_t|^2\|\bsg\|^2+\sum_{m=1}^{M}\ln\bigg(\alpha_{0,m,t}I_0(2\sqrt{\rho}|B_m[t]|)I_0(2\sqrt{\rho}|c_m[t] g_m|)\\
&+2\sum_{p=1}^{\infty}\alpha_{p,m,t}I_p(2\sqrt{\rho}|B_m[t]|)I_p(2\sqrt{\rho}|c_m[t] g_m|)\cos\left(p\left(\arg\left(B_m[t]\right)-\arg\left(c_m[t] g_m\right)\right)\right)\bigg).\nonumber
\end{align}

\subsection{'Better-than-optimal' Detector: Synchronous Operation}\label{subsec:SynchronousCC}

The system model in this case follows trivially from \eqref{eq:ReceivedTrainingCCNS} and \eqref{eq:ReceivedDataCCNS} by setting $\theta_m\equiv\theta$ and $\phi_m[t]\equiv\phi_t$, $\forall m\in\{1,\ldots,M\}$. Hence, the system model at time $t$ is given by
\begin{align}
\bsx&=\sqrt{\rho}e^{j\theta}\bsg + \bsw,\label{eq:ReceivedTrainingCCS}\\
\bsy_t&=\sqrt{\rho}e^{j\left(\theta_t+\phi_t\right)}\bsg s_t + \bsz_t,\hspace{5mm} t = 1,\ldots,\sfT,\label{eq:ReceivedDataCCS}
\end{align}
where $\theta_t\Define\theta+\sum_{\tau=1}^{t-1}\phi_\tau$. The likelihood for the causal ML symbol-by-symbol (SBS) detector for the symbol $s_t$ is given by
\begin{align}\label{eq:MLcausalSBSLikelihood}
p(\bsx,\bsy_1^t|s_t)=\sum_{\substack{s_\tau\in\mathcal{S},\\ \tau\in{1,\ldots,t-1}}}\idotsint_{-\pi}^\pi p(\bsx,\bsy_1^t|\theta_0,\phi_1^t,s_1^t)p(\theta_0)\prod_{\tau=1}^t p(\phi_\tau)\diff\theta_0\diff\phi_1^t.
\end{align}
The derivation of a closed-form expression of the above decision rule appears to be mathematically intractable. Hence, we are interested in deriving a genie-aided, 'better-than-optimal' detector, that will yield an optimistic performance bound on the detector in \eqref{eq:MLcausalSBSLikelihood}. Assume that before the detection of the symbol $s_t$, a genie provides the receiver with the exact knowledge of $\theta_t$. Then, the likelihood for ML causal SBS detector for $s_t$ is
\begin{align}\label{eq:GenieAidedConstantChannelLikelihood}
p(\bsy_t|\theta_t,s_t)=\int_{-\pi}^\pi p(\bsy_t|\theta_t,\phi_t,s_t)p(\phi_t)\diff\phi_t,
\end{align}
where
\begin{align}\label{eq:ConditionalDensityConstantChannel}
p(\bsy_t|\theta_t,\phi_t,s_t)&=\frac{1}{\pi^M}e^{-\left\|\bsy_t-\sqrt{\rho}e^{j\left(\theta_t+\phi_t\right)}\bsg s_t\right\|^2}\\
&=\frac{e^{-\left\|\bsy_t\right\|^2-\rho|s_t|^2\left\|\bsg\right\|^2}}{\pi^M}e^{2\sqrt{\rho}\left|\bsy_t^H\bsg s_t\right|\cos\left(\phi_t+\theta_t+\arg\left(s_t \bsy_t^H\bsg\right)\right)}.\nonumber
\end{align}
From \eqref{eq:PhaseNoiseIncrementsPDF} and \eqref{eq:ConditionalDensityConstantChannel} the likelihood in \eqref{eq:GenieAidedConstantChannelLikelihood} is given by
\begin{align}\label{eq:GenieAidedConstantChannelDetectorClosedFormExpression}
p(\bsy_t|\theta_t,s_t)&=\frac{e^{-\left\|\bsy_t\right\|^2-\rho|s_t|^2\left\|\bsg\right\|^2}}{\pi^M}\left(\beta_0+2\sum_{p=1}^{\infty}\beta_p\cos\left(p\left(\theta_t+\arg\left(s_t \bsy_t^H\bsg\right)\right)\right)\right),
\end{align}
where $\beta_p\Define\alpha_pI_p\left(2\sqrt{\rho}\left|\bsy_t^H\bsg s_t\right|\right)$. Finally the detector is given by
\begin{align}\label{eq:GenieAidedConstantChannelDetector}
\hat s_t=\argmax_{s_\tau\in\mathcal{S}}\hspace{3mm}-\rho|s_t|^2\left\|\bsg\right\|^2+\ln\left(\beta_0+2\sum_{p=1}^{\infty}\beta_p\cos\left(p\left(\theta_t+\arg\left(s_t \bsy_t^H\bsg\right)\right)\right)\right).
\end{align}


\subsection{Decision Feedback Detector: Non-Synchronous Operation}\label{subsec:NonSynchronousFC}

The system model for the fading channel case is given by
\begin{align}
\bsx&=\sqrt{\rho}\bsh + \bsw,\label{eq:ReceivedTrainingFCNS}\\
\bsy_t&=\sqrt{\rho}\bsPhi_t\bsh s_t + \bsz_t,\hspace{5mm} t = 1,\ldots,\sfT,\label{eq:ReceivedDataFCNS}
\end{align}
where the difference with \eqref{eq:ReceivedTrainingCCNS} and \eqref{eq:ReceivedDataCCNS} is the fading channel $\bsh\sim\mathcal{N}_\mathcal{C}(\bsO,\bsI_M)$. The initial phase reference $\bsTheta$ has been absorbed into $\bsh$ without modifying its statistics. The causal SBS decision feedback detector is derived also similarly to the constant channel case by assuming the following auxiliary model
\begin{align*}
\bsx&=\sqrt{\rho}\bsh+\bsw,\\
\bsy_\tau&=\sqrt{\rho}\bsD_\tau\bsh\hat s_\tau+\bsz_\tau,\hspace{5mm} \tau=1,\ldots,t-1\\
\bsy_t&=\sqrt{\rho}\bsPhi_t\bsh s_t+\bsz_t.
\end{align*}
The decision rule for this detector is given by \eqref{eq:DFReceiverCCNS}. The density $p(\bsx,\bsy_1^t|\hat s_1^{t-1},s_t)$ in this case is given by
\begin{align*}
p(\bsx,\bsy_1^t|\hat s_1^{t-1},s_t)&=\idotsint p(\bsx,\bsy_1^t|\hat s_1^{t-1},s_t,\bsPhi_t,\bsh)p(\bsh)p(\bsPhi_t)\diff\bsh\diff\bsPhi_t\\
&=\prod_{m=1}^{M}\iint p\left(x_m,\left\{y_m[\tau]\right\}_1^t\middle|\hat s_1^{t-1},s_t,\varphi_m[t],h_m\right)p(h_m)p(\varphi_m[t])\diff\varphi_m[t]\diff h_m.
\end{align*}
At the first step of the derivation the channel $h_m$ is marginalized out which yields the result for the density $p\left(x_m,\left\{y_m[\tau]\right\}_1^t\middle|\hat s_1^{t-1},s_t,\varphi_m[t]\right)$
\begin{align*}
&p\left(x_m,\left\{y_m[\tau]\right\}_1^t\middle|\hat s_1^{t-1},s_t,\varphi_m[t]\right)=\frac{1}{\pi^{t+1}\left(\hat a_m[t]+\rho|s_t|^2\right)}e^{-|x_m|^2-\sum_{\tau=1}^{t}|y_m[\tau]|^2+\frac{\rho\left(|\hat v_m[t]|^2+|y_m[t]|^2|s_t|^2\right)}{\hat a_m[t]+\rho|s_t|^2}}\\
&\cdot\left(I_0\left(\frac{2\rho|\hat \chi_m[t]|}{\hat a_m[t]+\rho|s_t|^2}\right)+2\sum_{p=1}^{\infty}I_p\left(\frac{2\rho|\hat \chi_m[t]|}{\hat a_m[t]+\rho|s_t|^2}\right)\cos\left(p\left(\varphi_m[t]+\arg\left(\hat \chi_m[t]\right)\right)\right)\right),
\end{align*}
where $\hat a_m[t]=1+\rho+\rho\sum_{\tau=1}^{t-1}|d_m[\tau]\hat s_\tau|^2$, $\hat v_m[t]\Define x_m^*+\sum_{\tau=1}^{t-1}y_m^*[\tau]d_m[\tau]\hat s_\tau$ and $\hat \chi_m[t]\Define\hat v_m^*[t]y_m^*[t]s_t$. In the following, the variable $\varphi_m[t]$ is marginalized out.
\begin{align*}
&p\left(x_m,\left\{y_m[\tau]\right\}_1^t\middle|\hat s_1^{t-1},s_t\right)=\int p\left(x_m,\left\{y_m[\tau]\right\}_1^t\middle|\hat s_1^{t-1},s_t,\varphi_m[t]\right)p(\varphi_m[t])\diff\varphi_m[t]\\
&=\frac{1}{\pi^{t+1}\left(\hat a_m[t]+\rho|s_t|^2\right)}e^{-|x_m|^2-\sum_{\tau=1}^{t}|y_m[\tau]|^2+\frac{\rho\left(|\hat v_m[t]|^2+|y_m[t]|^2|s_t|^2\right)}{\hat a_m[t]+\rho|s_t|^2}}\\
&\times\bigg(\alpha_{0,m,t}I_0\left(\frac{2\rho|\hat \chi_m[t]|}{\hat a_m[t]+\rho|s_t|^2}\right)+2\sum_{p=1}^{\infty}\alpha_{p,m,t}I_p\left(\frac{2\rho|\hat \chi_m[t]|}{\hat a_m[t]+\rho|s_t|^2}\right)\cos\left(p\left(\arg\left(\hat\chi_m[t]\right)\right)\right)\bigg).
\end{align*}
Finally, detector is given by
\begin{align}\label{eq:DecisionFeedbackFCNS}
\hat s_t&=\argmax_{s\in\mathcal{S}}\hspace{5mm}-\sum_{m=1}^{M}\ln\left( \hat a_m[t]+\rho|s_t|^2\right)+\sum_{m=1}^{M}\frac{\rho\left(|\hat v_m[t]|^2+|y_m[t]|^2|s_t|^2\right)}{\hat a_m[t]+\rho|s_t|^2}\\
&+\sum_{m=1}^{M}\ln\bigg(\alpha_{0,m,t}I_0\left(\frac{2\rho|\hat \chi_m[t]|}{\hat a_t+\rho|s_t|^2}\right)+2\sum_{p=1}^{\infty}\alpha_{p,m,t}I_p\left(\frac{2\rho|\hat \chi_m[t]|}{\hat a_t+\rho|s_t|^2}\right)\cos\left(p\left(\arg\left(\hat\chi_m[t]\right)\right)\right)\bigg).\nonumber
\end{align}

\subsection{'Better-than-optimal' Detector: Synchronous Operation}\label{subsec:SynchronousFC}

The system model is given by
\begin{align}\label{eq:FadingChannelSystemModel}
\bsx&=\sqrt{\rho}\bsh + \bsw,\\
\bsy_t&=\sqrt{\rho}e^{j\left(\sum_{\tau=1}^{t-1}\phi_\tau+\phi_t\right)}\bsh s_t + \bsz_t,\hspace{5mm} t = 1,\ldots,\sfT.
\end{align}
The likelihood for the causal ML symbol-by-symbol (SBS) detector for the symbol $s_t$ is given by
\begin{align}\label{eq:MLcausalSBSLikelihoodFC}
p(\bsx,\bsy_1^t|s_t)=\sum_{\substack{s_\tau\in\mathcal{S},\\ \tau\in{1,\ldots,t-1}}}\idotsint_{-\pi}^\pi p(\bsx,\bsy_1^t|\bsh,\phi_1^t,s_1^t)p(\bsh)\prod_{\tau=1}^t p(\phi_\tau)\diff\bsh\diff\phi_1^t.
\end{align}
We derive a genie-aided, 'better-than-optimal' detector, that will yield an optimistic performance bound on the detector in \eqref{eq:MLcausalSBSLikelihoodFC}. Assume that before the detection of the symbol $s_t$, a genie provides the receiver with the exact knowledge of $\sum_{\tau=1}^{t-1}\phi_\tau$ and the previously transmitted symbols, $s_1^{t-1}$. Then, the likelihood for ML SBS detector for $s_t$ is
\begin{align}\label{eq:GenieAidedFadingChannelLikelihood}
p(\bsx,\bsy_1^t|\phi_1^{t-1},s_1^{t-1},s_t)&=\int_{-\pi}^\pi\int_{\mathbb{C}^M} p(\bsx,\bsy_1^t|\bsh,\phi_1^{t-1},\phi_t,s_1^{t-1},s_t)p(\bsh)p(\phi_t)\diff\bsh\diff\phi_t
\end{align}
The conditional density $p(\bsx,\bsy_1^t|\bsh,\phi_1^{t-1},\phi_t,s_1^{t-1},s_t)$ is given by
\begin{align*}
p(\bsx,\bsy_1^t|\bsh,\phi_1^{t-1},\phi_t,s_1^{t-1},s_t)&=\frac{1}{\pi^{M(t+1)}}e^{-\left\|\bsx-\sqrt{\rho}\bsh\right\|^2-\sum_{\tau=1}^{t}\left\|\bsy_\tau-\sqrt{\rho}\bsh e^{j\left(\sum_{\hat\tau=1}^{\tau-1}\phi_{\hat \tau}+\phi_\tau\right)}s_\tau\right\|^2}.
\end{align*}
Define $a_t\Define 1+\rho+\rho\sum_{\tau=1}^{t-1}|s_\tau|^2$ and $\bsv_t\Define\bsx+\sum_{\tau=1}^{t}s_\tau^*e^{-j(\sum_{\hat \tau=1}^{\tau}\phi_{\hat \tau})}\bsy_\tau$.
The density $p(\bsx,\bsy_1^t|\phi_1^{t-1},\phi_t,s_1^{t-1},s_t)$ follows directly
\begin{align*}
&p(\bsx,\bsy_1^t|\phi_1^{t-1},\phi_t,s_1^{t-1},s_t)=\int_{\mathbb{C}^M} p(\bsx,\bsy_1^t|\bsh,\phi_1^{t-1},\phi_t,s_1^{t-1},s_t)p(\bsh)\diff\bsh\\
&=\frac{e^{-\left\|\bsx\right\|^2-\sum_{\tau=1}^t\left\|\bsy_\tau\right\|^2+\frac{\rho\left(\left\|\bsv_{t-1}\right\|^2+|s_t|^2\left\|\bsy_t\right\|^2\right)}{a_t+\rho|s_t|^2}+\frac{2\rho|s_t^*\bsv_{t-1}^H\bsy_t|\cos\left(\phi_t+\sum_{\tau=1}^{t-1}\phi_\tau-\arg\left(s_t^*\bsv_{t-1}^H\bsy_t\right)\right)}{a_t+\rho|s_t|^2}}}{\left(\pi^{t+1}\left(a_t+\rho|s_t|^2\right)\right)^M}.
\end{align*}
By further marginalization of $\phi_t$ and with the definitions
\begin{align*}
A_1&\Define\frac{e^{-\left\|\bsx\right\|^2-\sum_{\tau=1}^t\left\|\bsy_\tau\right\|^2+\frac{\rho\left(\left\|\bsv_{t-1}\right\|^2+|s_t|^2\left\|\bsy_t\right\|^2\right)}{a_t+\rho|s_t|^2}}}{\left(\pi^{t+1}\left(a_t+\rho|s_t|^2\right)\right)^M},\\
A_2&\Define\frac{2\rho|s_t^*\bsv_{t-1}^H\bsy_t|}{a_t+\rho|s_t|^2},\\
A_3&\Define\sum_{\tau=1}^{t-1}\phi_\tau-\arg\left(s_t^*\bsv_{t-1}^H\bsy_t\right),
\end{align*}
the likelihood $p(\bsx,\bsy_1^t|\phi_1^{t-1},\phi_t,s_1^{t-1},s_t)$ is given by
\begin{align}\label{eq:LikelihoodFadingChannelSynchronous}
p(\bsx,\bsy_1^t|\phi_1^{t-1},\phi_t,s_1^{t-1},s_t)&=A_1\left(\alpha_0I_0\left(A_2\right)+2\sum_{p=1}^{\infty}\alpha_pI_p\left(A_2\right)\cos\left(pA_3\right)\right).
\end{align}
Finally the detection rule is
\begin{align}\label{eq:GenieAidedFCS}
\hat s_t&=\argmax_{s_t\in\mathcal{S}}\hspace{5mm} \frac{\rho\left(\left\|\bsv_{t-1}\right\|^2+|s_t|^2\left\|\bsy_t\right\|^2\right)}{a_t+\rho|s_t|^2}-M\ln\left(a_t+\rho|s_t|^2\right)\\
&+\ln\left(\alpha_0I_0\left(\frac{2\rho|\chi[t]|}{a_t+\rho|s_t|^2}\right)+2\sum_{p=1}^{\infty}\alpha_pI_p\left(\frac{2\rho|\chi[t]|}{a_t+\rho|s_t|^2}\right)\cos\left(p\left(\sum_{\tau=1}^{t-1}\phi_\tau-\arg\left(\chi[t]\right)\right)\right)\right),\nonumber
\end{align}
where $\chi[t]\Define s_t^*\bsv_{t-1}^H\bsy_t$.

\subsection{Numerical Examples}

\begin{figure}[t!]
\centering
\begin{minipage}[t]{0.48\linewidth}
\includegraphics[width=\textwidth]{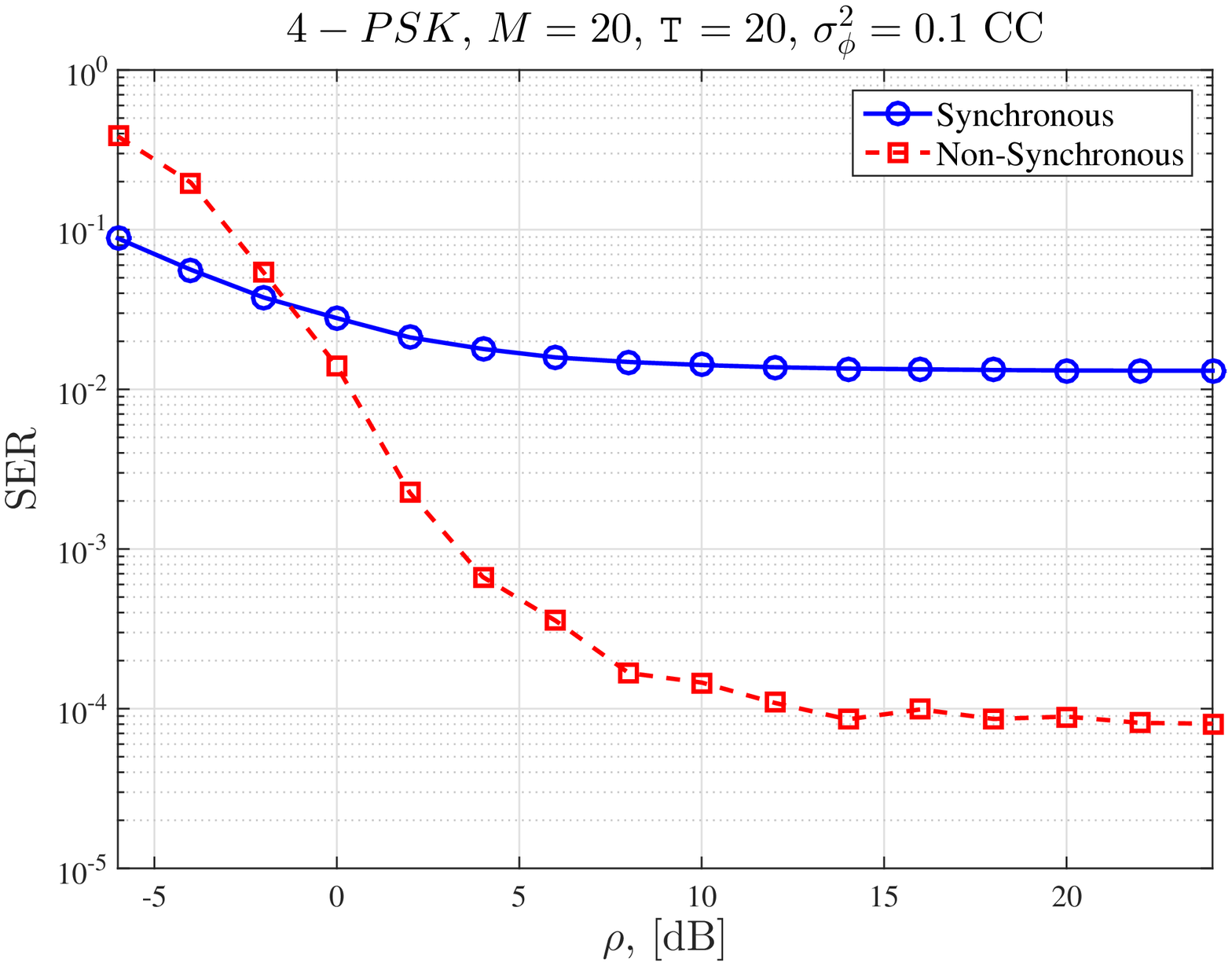}
\caption{Uncoded SER vs $\rho$ for QPSK symbols, $\sfT=20$, $M=20$ and $\sigma_\phi^2=0.1$ for the constant channel.}\label{fig:ConstantChannel20M20slotsQPSK0_1phN}
\end{minipage}
\quad
\begin{minipage}[t]{0.48\linewidth}
\includegraphics[width=\textwidth]{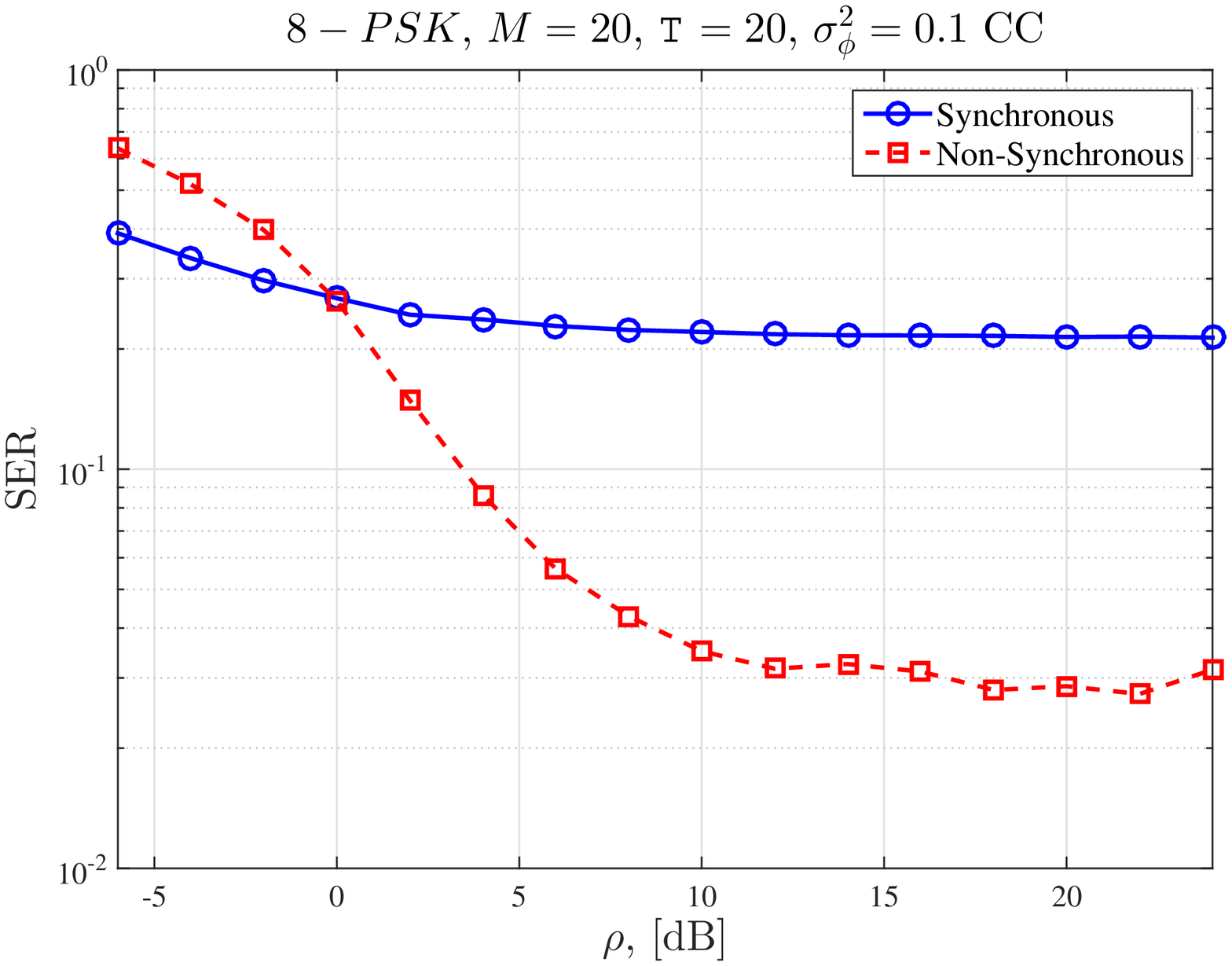}
\caption{Uncoded SER vs $\rho$ for 8-PSK symbols, $\sfT=20$, $M=20$ and $\sigma_\phi^2=0.1$ for the constant channel.}\label{fig:ConstantChannel20slots8PSK0_1phN}
\end{minipage}
\end{figure}

\begin{figure}[t!]
\centering
\begin{minipage}[t]{0.48\linewidth}
\includegraphics[width=\textwidth]{FadingChannel20slotsQPSK0_07phN}
\caption{Uncoded SER vs $\rho$ for QPSK symbols, $\sfT=20$, $M=20$ and $\sigma_\phi^2=0.07$ for the fading channel.}\label{fig:FadingChannel20slotsQPSK0_07phN}
\end{minipage}
\quad
\begin{minipage}[t]{0.48\linewidth}
\includegraphics[width=\textwidth]{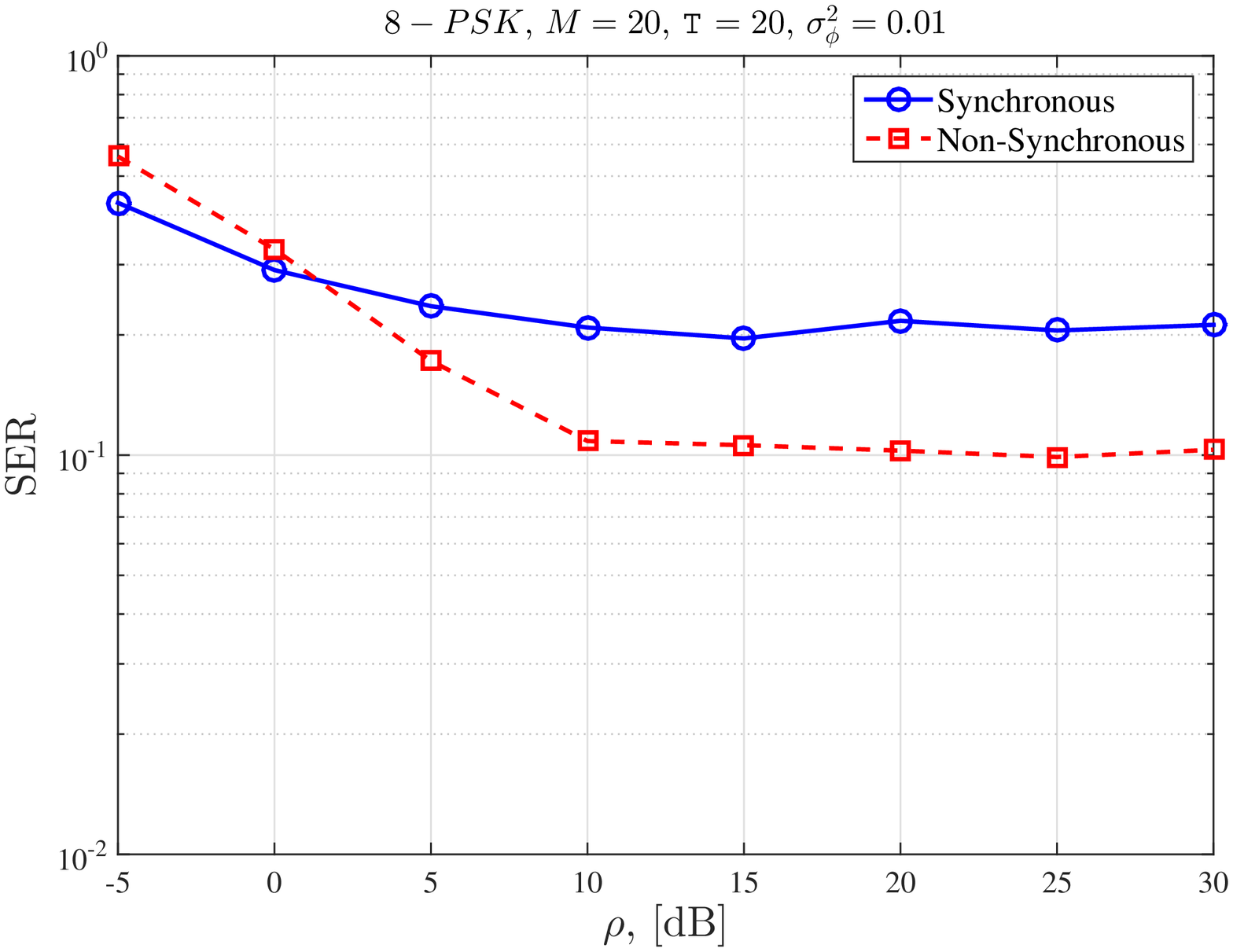}
\caption{Uncoded SER vs $\rho$ for 8-PSK symbols, $\sfT=20$, $M=20$ and $\sigma_\phi^2=0.01$ for the fading channel.}\label{fig:FadingChannel20slots8PSK0_01phNv2}
\end{minipage}
\end{figure}

In this section, we provide some numerical examples to investigate the performance of the various derived detectors. In all the examples, the phase noise increments $\phi_m[t],~m=1,\ldots,M,~t=1,\ldots,\sfT$ are assumed to be i.i.d. zero mean wrapped Gaussian random variables with variance $\sigma_\phi^2$. Hence, the pdf of the increments is given by \eqref{eq:PhaseNoiseIncrementsPDF} with $\alpha_{m,p}=\exp\left(-\frac{\sigma_\phi^2}{2}\right)^{p^2}$. Also in all the Figures $M=20$ and the length of the data interval is $\sfT=20$. In Fig. \ref{fig:ConstantChannel20M20slotsQPSK0_1phN} the performance of the genie-aided receiver for the synchronous operation in \eqref{eq:GenieAidedConstantChannelDetector} is compared with the suboptimal decision-feedback detector for the non-synchronous operation in \eqref{eq:DecisionFeedbackCCNS} for the constant channel case. The input symbols are selected from a QPSK constellation with equal probability and the variance of the phase noise increments is $\sigma_\phi^2=0.1$. In Fig. \ref{fig:ConstantChannel20slots8PSK0_1phN} the uncoded SER performance of the detectors \eqref{eq:GenieAidedConstantChannelDetector} and \eqref{eq:DecisionFeedbackCCNS} is shown as a function of $\rho$ for equiprobable 8-PSK symbols. In Figs. \ref{fig:FadingChannel20slotsQPSK0_07phN} and \ref{fig:FadingChannel20slots8PSK0_01phNv2} the uncoded SER performance of the detectors in \eqref{eq:DecisionFeedbackFCNS} and \eqref{eq:GenieAidedFCS} is plotted as a function of $\rho$. In Fig. \ref{fig:FadingChannel20slotsQPSK0_07phN} the symbols are QPSK with equal probability and $\sigma_\phi^2=0.07$, and in Fig. \ref{fig:FadingChannel20slots8PSK0_01phNv2} the symbols are equiprobable 8-PSK symbols with $\sigma_\phi^2=0.01$. In all cases shown in Figs. \ref{fig:ConstantChannel20M20slotsQPSK0_1phN}-\ref{fig:FadingChannel20slots8PSK0_01phNv2}, it is clear that the decision-feedback non-synchronous detector outperforms the corresponding genie-aided synchronous detector. This establishes the validity of the results to setups with more data channel uses.


\bibliographystyle{ieeetr}
\bibliography{phNbib}

\end{document}